\pdfoutput=1
\documentclass[cernpreprint,atlasdraft=false,texlive=2016,UKenglish,texmf]{atlasdoc}
 
\usepackage[subfigure]{atlaspackage}
\usepackage[backref=false]{atlasbiblatex}

\usepackage{atlasphysics}
\usepackage{atlasbsm}
\usepackage{atlasjetetmiss}
\usepackage{atlasprocess}
 
\usepackage{pifont}
\usepackage{dcolumn}
\usepackage{xcolor,colortbl}
\usepackage{float}

\definecolor{Gray}{gray}{0.85}
\definecolor{LightCyan}{rgb}{0.88,1,1}
\definecolor{cZjets}{rgb}{0.9882, 0.6902, 0.03137}
\definecolor{cWjets}{rgb}{0.5098, 0.3725,0.5294}
\definecolor{cTtbar}{rgb}{0.48235, 0.6980, 0.4549}
\definecolor{cSt}{rgb}{0.21176,0.47451,0.74901}
\definecolor{cOther}{rgb}{0.65882, 0.64313, 0.588235}

\newcolumntype{C}{>{\columncolor{green!50}}c}
\newcolumntype{O}{>{\columncolor{orange!30}}c}
\newcolumntype{Y}{>{\columncolor{yellow!30}}c}
\newcolumntype{a}{>{\columncolor{Gray}}c}
\newcolumntype{b}{>{\columncolor{white}}c}

\graphicspath{{logos/}{figures/}}
 
\usepackage{ANA-SUSY-2018-34-PAPER-defs}

\AtlasTitle{Search for new phenomena in final states with $b$-jets and missing transverse momentum in $\sqrt{s}=13~\mathrm{TeV}$ \(pp\) collisions with the ATLAS detector}
 
\AtlasJournal{JHEP}
\AtlasAbstract{
The results of a search for new phenomena in final states with $b$-jets and missing transverse momentum using 139~fb$^{-1}$ of proton--proton data collected at a centre-of-mass energy $\sqrt{s} = 13$~TeV by the ATLAS detector at the LHC are reported. The analysis targets final states produced by the decay of a pair-produced supersymmetric bottom squark into a bottom quark and a stable neutralino. The analysis also seeks evidence for models of pair production of dark matter particles produced through the decay of a generic scalar or pseudoscalar mediator state in association with a pair of bottom quarks, and models of pair production of scalar third-generation down-type leptoquarks. No significant excess of events over the Standard Model background expectation is observed in any of the signal regions considered by the analysis. Bottom squark masses below \limitLargeSplitSbottom are excluded at 95\% confidence level if the neutralino is massless. In the case of nearly mass-degenerate bottom squarks and neutralinos, the use of dedicated secondary-vertex identification techniques permits the exclusion of bottom squarks with masses up to \limitSmallSplitSbottom for mass splittings between the squark and the neutralino of \SplitLimit. These limits extend substantially beyond the regions of parameter space excluded by similar ATLAS searches performed previously.
}
 
\author{The ATLAS Collaboration}
 
\AtlasRefCode{SUSY-2018-34}
 
\PreprintIdNumber{CERN-EP-2021-001}
 
\AtlasJournalRef{JHEP 05 (2021) 093}
\AtlasDOI{10.1007/JHEP05(2021)093}

\hypersetup{pdftitle={ATLAS document},pdfauthor={The ATLAS Collaboration}}
 
\begin{document}
 
\maketitle

\newpage
 
\section{Introduction}
 
The possible existence of non-luminous matter in the universe, referred to as dark matter (DM), is supported by a wide variety of astrophysical and cosmological measurements~\cite{Hinshaw:2012aka,Akrami:2018vks,Trimble1987,Bertone:2004pz,Feng:2010gw}. However, the nature and properties of the DM remain largely unknown and represent one of the most important unanswered questions in physics. A plausible candidate for cold dark matter~\cite{Goldberg:1983nd,Ellis:1983ew} is the stable lightest neutralino ($\ninoone$) in $R$-parity-conserving models~\cite{Farrar:1978xj} of electroweak scale supersymmetry (SUSY)~\cite{Golfand:1971iw,Volkov:1973ix,Wess:1974tw,Wess:1974jb,Ferrara:1974pu,Salam:1974ig}. In supersymmetric models that naturally address the gauge hierarchy problem~\cite{Sakai:1981gr,Dimopoulos:1981yj,Ibanez:1981yh,Dimopoulos:1981zb}, the scalar partners of the third-generation quarks are light~\cite{Barbieri:1987fn,deCarlos:1993yy}. This may lead to the lighter bottom squark (\sbottomone) and top squark (\stopone) mass eigenstates\footnote{The scalar partners of the left-handed and right-handed chiral components of the bottom quark (\sbottomL, \sbottomR) or top quark (\stopL, \stopR) mix to form two mass eigenstates in each case, of which the \sbottomone and the \stopone are defined to be the lighter.} being significantly lighter than the other squarks and gluinos. As a consequence, the \sbottomone and \stopone could be pair produced with relatively large cross-sections in $pp$ collisions at the Large Hadron Collider (LHC~\cite{Evans:2008zzb}). In most SUSY models, the \sbottomone\ and the \stopone decay into final states incorporating third-generation quarks and invisible $\ninoone$ particles.
 
More generically, the dark matter may be composed of weakly interacting massive particles (WIMPs, generically denoted by $\chi$ in the rest of the paper)~\cite{Steigman:1984ac}, of which the lightest supersymmetric particle (LSP) is one example. WIMPs can account for the measured relic density of dark matter in the early universe across a broad portion of parameter space~\cite{Kolb:1990vq,Akrami:2018vks,Hinshaw:2012aka}. WIMPs could be produced in pairs at the LHC through the decay of a new mediator particle coupling to Standard Model (SM) quarks \cite{Fox:2008kb,Cassel:2009pu,Bai:2010hh,Abdallah:2015ter,Abercrombie:2015wmb,2HDMWGproxi}. Should this mediator preferentially couple to third-generation quarks then an excess of events containing such quarks along with invisible dark matter particles could be observed. Such events can be described in the framework of simplified DM models~\cite{Abercrombie:2015wmb,Buckley:2014fba,Haisch:2015ioa} with model assumptions described in Refs.~\cite{Abercrombie:2015wmb,Boveia:2016mrp,Albert:2017onk,2HDMWGproxi}.
 
\begin{figure}[htpb]
\begin{center}
\subfigure[]{\includegraphics[width=0.29\textwidth]{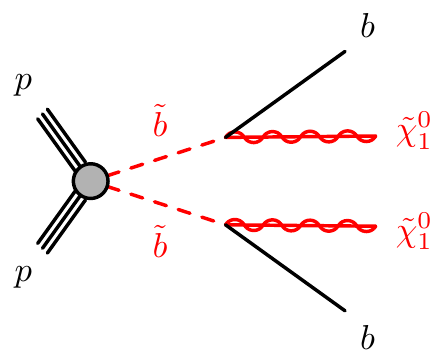}\label{fig:bbMETdiagram1}}\hfill
\subfigure[]{\includegraphics[width=0.29\textwidth]{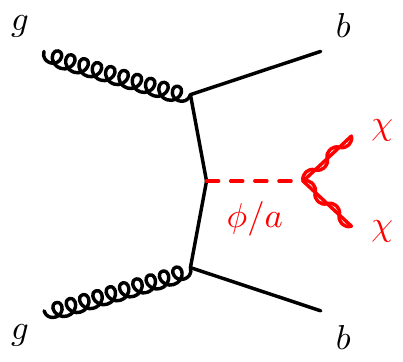}\label{fig:bbMETdiagram2}}\hfill
\subfigure[]{\includegraphics[width=0.29\textwidth]{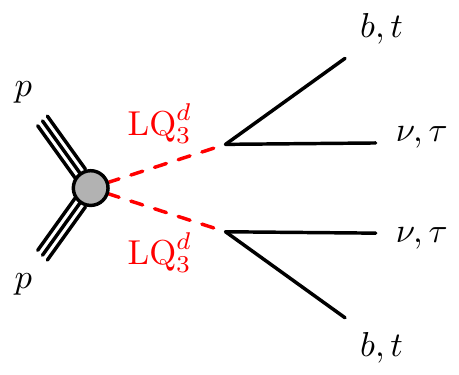}\label{fig:bbMETdiagram3}}
\end{center}
\caption{Diagrams illustrating the processes targeted by this analysis: (a) bottom squark pair production, (b) production of DM particles (indicated with $\chi$) through the decay of a scalar or pseudoscalar mediator coupling to bottom quarks, and (c) pair production of scalar third-generation down-type leptoquarks decaying to bottom quarks and neutrinos or top quarks and $\tau$-leptons. BSM particles are indicated in red, while SM particles are indicated in black.
}
\end{figure}

This paper describes a search for the production of invisible dark matter particles in association with bottom quarks. Signal regions (SRs) are developed which target the direct pair production of bottom squarks, each of which decays into a \ninoone and a bottom quark, as shown in Figure~\ref{fig:bbMETdiagram1}. Additional signal regions target the pair production of DM particles through the decay of a generic scalar ($\phi$) or pseudoscalar ($a$) mediator state produced in association with a pair of bottom quarks (Figure~\ref{fig:bbMETdiagram2}). The results of the analysis are also interpreted in the context of beyond-the-SM (BSM) scenarios incorporating pair-produced scalar third-generation down-type leptoquarks LQ$^d_3$ \cite{dimop_suss,techni2,techni3,string,comp,pati_salam_colour,georgi_glashow_unification,Buchmuller:1986zs} decaying to bottom quarks and neutrinos or top quarks and $\tau$-leptons (Figure~\ref{fig:bbMETdiagram3}). These models are all characterised by events consisting of jets containing $b$-hadrons (referred to as \bjets), missing transverse momentum (\met), and no charged leptons.
 
Previous searches by ATLAS \cite{Aad:2015pfx,Aaboud:2017wqg,SUSY-2016-18,Aaboud:2019bye} and CMS \cite{CMS-SUS-19-005,Sirunyan:2019ctn} using comparable or smaller datasets have targeted similar final states. This analysis extends the regions of parameter space probed by the LHC through the use of a larger dataset than in previous ATLAS searches, new boosted decision tree (BDT) discriminants, and also new selections maximising the efficiency for reconstructing \bjets with low transverse momentum generated by, for instance, SUSY models with small mass-splitting between \sbottomone and \ninoone.
 
Section~\ref{sec:detector} presents a brief overview of the ATLAS detector, Section~\ref{sec:dataMC} describes the data and simulation samples used in the analysis and Section~\ref{sec:reconstruction} presents the methods used to reconstruct events. An overview of the analysis strategy, including background estimation, is presented in Section~\ref{sec:analysis_strategy}. The systematic uncertainties considered in the analysis are described in Section~\ref{sec:systematic_uncertainties}. Section~\ref{sec:results_and_interpretation} presents the results and interpretation thereof. The conclusions of the analysis are presented in Section~\ref{sec:conclusion}.
 
\FloatBarrier


\section{ATLAS detector}
\label{sec:detector}
The ATLAS  detector~\cite{Aad:2008zzm,B-layerRef,PIX-2018-001} is a      multipurpose detector with a forward--backward symmetric cylindrical
geometry and nearly 4$\pi$ coverage in solid angle.\footnote{ATLAS uses a right-handed coordinate system with its origin at the nominal interaction point in the centre of the detector. The positive $x$-axis is defined by the direction from the interaction point to the centre of the LHC ring, with the positive $y$-axis pointing upwards, while the beam direction defines the $z$-axis. Cylindrical coordinates $(r,\phi)$ are used in the transverse plane, $\phi$ being the azimuthal angle around the $z$-axis. The transverse momentum $\pt$, the transverse energy $\ET$ and the missing transverse momentum are defined in the $x$--$y$ plane unless stated otherwise. The pseudorapidity $\eta$ is defined in terms of the polar angle $\theta$ by $\eta=-\ln\tan(\theta/2)$ and the rapidity is defined as $y = (1/2)\ln[(E+p_z)/(E-p_z)]$ where $E$ is the energy and $p_z$ the longitudinal momentum of the object of interest.} The inner detector (ID) tracking system consists of pixel and silicon microstrip detectors covering the pseudorapidity region $|\eta|<2.5$, surrounded by a transition radiation tracker, which improves electron identification over the region $|\eta|<2.0$.
The ID is surrounded by a thin superconducting solenoid providing an axial 2~T magnetic field and by a fine-granularity lead/liquid-argon (LAr) electromagnetic calorimeter covering $|\eta|<3.2$. A steel/scintillator-tile calorimeter provides hadronic coverage in the central pseudorapidity range ($|\eta|<1.7$). The endcap and forward calorimeters ($1.5<|\eta|<4.9$) are made of LAr active layers with either copper or tungsten as the absorber material for electromagnetic and hadronic measurements. The muon spectrometer with an air-core toroid magnet system surrounds the calorimeters. Three layers of high-precision tracking chambers provide coverage in the range $|\eta|<2.7$, while     dedicated chambers allow triggering in the region $|\eta|<2.4$.
 
\section{Data collection and simulated event samples}
\label{sec:dataMC}
 
The data analysed in this paper were collected between 2015 and 2018 at a centre-of-mass energy of 13~\TeV\ with a $25$~ns proton bunch crossing interval. The average number of \pp interactions per bunch crossing (pile-up) ranged from 13 in 2015 to around 38 in 2017--2018. Application of beam, detector and data-quality criteria \cite{Aad:2700249} results in a total integrated luminosity of 139~\ifb. The uncertainty in the combined 2015--2018 integrated luminosity is 1.7\% \cite{ATLAS-CONF-2019-021}, obtained using the LUCID-2 detector \cite{LUCID2} for the primary luminosity measurements and cross-checked by a suite of other systems.
 
Events are required to pass a missing transverse momentum trigger~\cite{ATLASTrigger2016Paper,TRIG-2019-01} with an online threshold of $70$--$110~\GeV$, depending on the data-taking period. This trigger is found \cite{TRIG-2019-01} to have an efficiency greater than 95\% for events satisfying the offline selections of the analysis. Additional single-lepton triggers requiring the presence of electrons or muons are used in the two-lepton control regions defined in Section~\ref{sec:analysis_strategy} to estimate the background originating from \Zjets production \cite{TRIG-2018-01,TRIG-2018-05}. These triggers yield an approximately constant efficiency in the presence of a single isolated electron or muon with transverse momentum (\pt) greater than 27~\GeV.
 
Monte Carlo (MC) simulations are used to model SM background processes and the SUSY, dark matter and leptoquark signals considered in the analysis. Samples of bottom squark and dark matter signal events were generated with \MGMCatNLOV{2.6.2}~\cite{Alwall:2014hca} at leading order (LO) in the strong coupling constant ($\alphas$), with the renormalisation and factorisation scales set to $\HTgen/2$ (where \HTgen is the scalar sum of the transverse momenta of the outgoing partons) and parton distribution function (PDF) NNPDF2.3 LO~\cite{Ball:2012cx}. The matrix element (ME) calculations were performed at tree level and include the emission of up to two additional partons. Bottom squarks decayed directly into a \ninoone and a bottom quark with 100\% branching ratio, as is the case in $R$-parity-conserving models in which the lighter bottom squark is the next-to-lightest supersymmetric particle. Leptoquark signal events were generated at next-to-leading order (NLO) in $\alphas$ with \MGMCatNLOV{2.6.0}~\cite{Alwall:2014hca}, using the leptoquark model of Ref.~\cite{Mandal:2015lca} that adds parton showers to previous fixed-order NLO QCD calculations~\cite{Kramer:2004df,Kramer:1997hh}, and the NNPDF3.0 NLO~\cite{Ball:2014uwa} PDF set with $\alphas=0.118$. In all cases, simulated signal events were passed to \PYTHIAV{8.230}~\cite{Sjostrand:2014zea} for parton showering (PS) and hadronisation. ME--PS matching was performed following the CKKW-L prescription~\cite{CKKW}, with a matching scale set to one quarter of the mass of the bottom squark or leptoquark.
 
Bottom squark pair-production cross-sections were calculated at approximate next-to-next-to-leading-order (NNLO) accuracy in $\alphas$, also adding contributions from the resummation of soft gluon emission at next-to-next-to-leading-logarithm accuracy (approximate NNLO+NNLL)~\cite{Beenakker:2016lwe,Beenakker:1997ut,Beenakker:2010nq,Beenakker:2016gmf}. The nominal cross-sections and their uncertainties were derived using the PDF4LHC15\_mc PDF set, following the recommendations of Ref.~\cite{Butterworth:2015oua}. For  \sbottomone masses ranging from 400~\GeV\ to 1.5~\TeV, the cross-sections range from 2.1\,pb to 0.26\,fb, with uncertainties ranging from 7\% to 17\%. Leptoquark signal cross-sections were obtained from the calculation of direct top squark pair production, as this process has the same production modes, computed at approximate next-to-next-to-leading order (NNLO) in $\alphas$ with resummation of next-to-next-to-leading logarithmic (NNLL) soft gluon terms~\cite{Beenakker:2016lwe,Beenakker:1997ut,Beenakker:2010nq,Beenakker:2016gmf}. The cross-sections do not include lepton $t$-channel contributions, which are neglected in Ref.~\cite{Mandal:2015lca} and may lead to corrections at the percent level~\cite{Borschensky:2020hot}.
 
The production cross-sections for generic scalar and pseudoscalar mediators were evaluated including NLO QCD corrections assuming SM Yukawa couplings to quarks, in a five-flavour scheme, following the prescriptions of Ref.~\cite{arXiv:1811.08002}. They were calculated with renormalisation and factorisation scales set to $\HTgen/3$ and the jet $\pT$ threshold (`ptj' in Ref.~\cite{arXiv:1811.08002}) set to 20~\GeV. They range from about 29~pb to about 1.5~fb for mediator masses between 10~\GeV\ and 500~\GeV.
 
The SM backgrounds considered in this analysis are: \Zjets production; \Wjets production;  \ttbar pair production; single-top-quark production; \ttbar production in association with electroweak or Higgs bosons ($t\bar{t}+X$); and diboson production ($WW$, $ZZ$, $ZW$, $ZH$ and $WH$). The events were simulated using different MC generator programs depending on the process. Details of the generators, PDF set and underlying-event tuned parameter set (tune) used for each process are listed in Table~\ref{tab:mc_samples}.
 
\begin{table}[htbp]
\centering
\caption{The SM background MC simulation samples used in this paper. Generator, PDF set, parton shower, tune used for the underlying event (UE), and order in $\alphas$ of cross-section calculations used for yield normalisation, are shown for each process considered.   }
\resizebox{\columnwidth}{!}{
\begin{tabular}{ l | | lllll }
\hline\hline
Process            & ME event generator                    & PDF       & PS and        & UE tune & Cross-section                                                                                                 \\
&                                       &              & hadronisation &         & calculation                                                                                                   \\
\hline\hline
$V$+jets $(V=W/Z)$ & \SHERPA~2.2.1~\cite{Gleisberg:2008ta} & NNPDF3.0 NNLO & \SHERPA       & Default & NNLO~\cite{Catani:2009sm}                                                                                     \\
$\ttbar$           & \POWHEGBOXV{v2}~\cite{Alioli:2010xd}       & NNPDF3.0 NNLO   & \PYTHIAV{8.230}  & A14     & NNLO+NNLL~\cite{Czakon:2013goa,Czakon:2012pz,Czakon:2012zr,Baernreuther:2012ws,Cacciari:2011hy,Czakon:2011xx} \\
Single top         & \POWHEGBOXV{v2}                            & NNPDF3.0 NNLO   & \PYTHIAV{8.230}  & A14     & NNLO+NNLL~\cite{Kidonakis:2011wy,Kidonakis:2010ux,Kidonakis:2010tc}                                           \\
Diboson            & \SHERPA~2.2.1--2.2.2                   & NNPDF3.0 NNLO & \SHERPA       & Default & NLO    \\
$\ttbar+V $        & \ \AMCatNLOV{2.3.3}                          & NNPDF3.0 NLO     & \PYTHIAV{8.210}  & A14     & NLO~\cite{Alwall:2014hca}                                                                                     \\
$\ttbar H$         & \ \AMCatNLOV{2.2.3}                          & NNPDF3.0 NLO   & \PYTHIAV{8.230}  & A14     & NLO~\cite{Beenakker:2002nc,Dawson:2003zu,Yu:2014cka,Frixione:2015zaa}   \\
\hline\hline
\end{tabular}
}
\label{tab:mc_samples}
\end{table}
 
The {\textsc EvtGen}~v1.6.0 program~\cite{Lange:2001uf} was used to describe the properties of the $b$- and $c$-hadron decays in the signal samples and in the background samples, except those produced with \SHERPA. For all SM background samples, the response of the detector to particles was modelled with the full ATLAS detector simulation \cite{:2010wqa} based on \textsc{Geant4} \cite{Agostinelli:2002hh}. Signal samples were prepared using a fast simulation based on a parameterisation of showers in the ATLAS electromagnetic and hadronic calorimeters \cite{ATL-PHYS-PUB-2010-013} coupled to \textsc{Geant4} simulations of particle interactions elsewhere. All simulated events were overlaid with multiple $pp$ collisions simulated with \PYTHIA~8.186 using the A3 tune~\cite{A14tune} and the NNPDF2.3 LO PDF set~\cite{Ball:2012cx}. The MC samples were generated with variable levels of pile-up in the same and neighbouring collisions, and were reweighted to match the distribution of the mean number of interactions observed in data in 2015--2018.


\section{Event reconstruction}
\label{sec:reconstruction}
 
The analysis identifies events with jets containing $b$-hadrons or secondary vertices corresponding to $b$-hadron decays, missing transverse momentum from the $\chi$ or $\ninoone$, and no charged leptons (electrons or muons). The last requirement is effective in suppressing SM backgrounds arising from $W\rightarrow\ell\nu$ decays, including events containing top quark production.
 
Events are required to have a primary vertex~\cite{PERF-2015-01,ATL-PHYS-PUB-2015-026} reconstructed from at least two tracks~\cite{PERF-2015-08} with $\pT>0.5~\GeV$. If more than one such vertex is found, the one with the largest sum of the squares of transverse momenta of associated tracks~\cite{ATL-PHYS-PUB-2015-026} is selected as the hard-scattering collision.
 
Jet candidates are reconstructed using the anti-$k_t$ jet algorithm~\cite{Cacciari:2008gp, Cacciari:2011ma} with radius parameter $R=0.4$~\cite{PERF-2014-07} using particle-flow objects (PFOs)~\cite{PERF-2015-09} as inputs. PFOs are charged-particle tracks matched to the hard-scatter vertex with the requirement $|z_0 \sin{\theta}|<2.0~$mm, where $z_0$ is the longitudinal impact parameter,\footnote{The transverse impact parameter is defined as the distance of closest approach of a track to the beam-line, measured in the transverse plane. The longitudinal impact parameter corresponds to the $z$-coordinate distance between the point along the track at which the transverse impact parameter is defined and the primary vertex.} and calorimeter energy clusters surviving an energy subtraction algorithm that removes the calorimeter deposits of good-quality tracks from any vertex. Jet energy scale corrections, derived from MC simulation and data, are used to calibrate the average energies of jet candidates to the scale of their constituent particles~\cite{JETM-2018-05}. Only corrected jet candidates with $\pT > 20$~\GeV~and $|\eta|<2.8$ are considered explicitly when selecting events in this analysis, although jet candidates lying within   $|\eta| \leq 4.5$ are considered when calculating $\met$. A set of quality criteria is applied to identify jets which arise from non-collision sources or detector noise~\cite{Aad:2013zwa} and any event which contains a jet failing to satisfy these criteria is removed. Jets containing a large particle momentum contribution from pile-up vertices, as measured by the jet vertex tagger (JVT) discriminant~\cite{PERF-2014-03} are rejected if they have \pT$\in [20,60]$~\GeV, $|\eta|<2.4$ and a discriminant value of $\text{JVT} < 0.5$.
 
Selected jets are identified as \bjets if they lie within the ID acceptance of $|\eta| < 2.5$ and are tagged by a multivariate algorithm (\verb|DL1r|) which uses a selection of inputs including information about the impact parameters of ID tracks, the presence of displaced secondary vertices and the reconstructed flight paths of $b$- and $c$-hadrons inside the jet~\cite{Aad:2019aic}. The \btagging algorithm uses a working point with an efficiency of 77\%, determined with a sample of simulated \ttbar events.  The corresponding misidentification (mis-tag) rate is 20\% for $c$-jets and 0.9\% for light-flavour jets. Differences in efficiency and mis-tag rate between data and MC simulation are taken into account with correction factors as described in Ref.~\cite{Aad:2019aic}.
 
To enhance sensitivity to models where low-\pt bottom quarks are present in the final state (e.g.\ bottom squark pair production with nearly mass-degenerate \sbottomone and \ninoone), a dedicated secondary-vertex finding algorithm (\softbvrt) is used. Documented in Ref.~\cite{ATLAS-CONF-2019-027}, this algorithm reconstructs secondary vertices independently of the presence of an associated jet. A new {\em loose} working point, defined using the same track and vertex variables described in Ref.~\cite{Aad:2020srt} for the {\em medium} and {\em tight} working points, was optimised for this analysis. The efficiency to correctly identify the secondary vertex associated with the decay of a $b$-hadron (\epsilonvtx) ranges from 5\% for a $b$-hadron \pt of 5~\GeV\ to 40\% for a \pt of 15~\GeV. The corresponding probability (\fvtx) to obtain a vertex in an event without a $b$-hadron depends on the event topology and pile-up conditions, and is 1\%--5\%. Differences in \epsilonvtx (\fvtx) between data and MC simulation are taken into account by using correction factors computed in dileptonic \ttbar (\Wjets) production events. The correction factors are compatible with one for \epsilonvtx and range between 1.2 and 1.5 for \fvtx.
 
Two different classes (`baseline' and `high-purity') of reconstructed lepton candidates (electrons or muons) are used in the analyses presented here. When selecting samples for the search, events containing a `baseline' electron or muon are rejected. When selecting events with leptons for the purpose of estimating $\Wjets$, $\Zjets$ and top quark backgrounds, additional requirements are applied to leptons to ensure greater purity of these backgrounds. These leptons are referred to as `high-purity' leptons in the following and form a subset of the baseline leptons.
 
Baseline muon candidates are formed by combining information from the muon spectrometer and ID as described in Refs.~\cite{PERF-2015-10,MUON-2018-03} and are required to possess $\pt > 6 ~\GeV$ and $|\eta| <2.7$.  Baseline muon candidates must additionally have a significance of the transverse impact parameter relative to the beam-line $|d_0^{\mathrm{BL}}|/\sigma(d_0^{\mathrm{BL}}) <3$, and a longitudinal impact parameter relative to the primary vertex $|z_0~\sin(\theta)|< 0.5$~mm. Furthermore, high-purity muon candidates must satisfy the {\em Medium} identification requirements described in Refs.~\cite{PERF-2015-10,MUON-2018-03} and the {\em FixedCutTightTrackOnly} isolation requirements, which are described in the same references and use tracking-based variables to implement a set of $\eta$- and $\pt$-dependent criteria.
 
Baseline electron candidates are reconstructed from an isolated electromagnetic calorimeter energy deposit matched to an ID track \cite{Aad:2019tso} and are required to possess $\pt > 7~\GeV$ and $|\eta| < 2.47$, and to satisfy the {\em Loose} likelihood-based identification criteria described in Refs.~\cite{Aad:2019tso,Aaboud:2019ynx}. High-purity electron candidates are also required to possess $|d_0^{\mathrm{BL}}|/\sigma(d_0^{\mathrm{BL}}) < 5$ and $|z_0 \sin(\theta)|<0.5$~mm, and to satisfy {\em Tight} isolation requirements~\cite{Aad:2019tso,Aaboud:2019ynx}.
 
High-purity muon and electron candidates used to estimate backgrounds in this analysis are required to possess $\pt > 20~\GeV$ in order to reduce the impact of misidentified or non-prompt leptons. In addition, when using events selected with single-lepton triggers, the leading lepton is required to possess $\pt > 27~\GeV$ in order to ensure that events are selected in the trigger plateau.
 
After the selections described above, a procedure is applied to remove non-isolated leptons and avoid double counting of tracks and energy depositions associated with overlapping reconstructed jets, electrons and muons. The procedure applies the following actions to the event. First, baseline electrons are discarded if they share an ID track with a baseline muon. Next, any jet with $|\eta|<2.8$ lying within a distance $\Delta R\equiv\sqrt{(\Delta y)^2+(\Delta\phi)^2}=0.2$ of a baseline electron is discarded and the electron is retained. Similarly, any jet with $|\eta|<2.8$ satisfying $N_\text{trk} < 3$ (where $N_\text{trk}$ refers to the number of tracks with $\pT>500~\MeV$ that are associated with the jet) within $\Delta R\equiv\sqrt{(\Delta y)^2+(\Delta\phi)^2}=0.2$ of a baseline muon is discarded and the muon is retained. Finally, baseline electrons or muons lying within a distance $\Delta R =\textrm{min}(0.4, 0.04 + 10~\GeV{}/p_{\textrm T}^{e/\mu})$ of a remaining jet are discarded.
 
Multiplicative scale factors are applied to simulated events to account for differences between data and simulation for the lepton trigger, reconstruction, identification and isolation efficiencies, and for the jet momentum scales and energy resolutions. Similar corrections are also applied to the probability of mis-tagging jets originating from the hard $pp$ scattering as pile-up jets with the JVT discriminant.
 
The missing transverse momentum \ptmiss, whose magnitude is referred to as  $\met$, is defined as the negative vector sum of the \pt of all selected and calibrated physics objects (electrons, muons, photons and jets) in the event, with an extra term added to account for energy in the event that is not associated with any of these objects~\cite{ATLAS-CONF-2018-023}. This last `soft term' contribution is calculated from the ID tracks with $\pT>500~\MeV$ associated with the primary vertex, thus ensuring that it is robust against pile-up contamination~\cite{PERF-2016-07,ATLAS-CONF-2018-023}. Photons contributing to the \ptmiss calculation are required to satisfy $\pt > 25~\GeV$ and $|\eta| < 2.37$ (excluding the transition region $1.37 < |\eta| < 1.52$ between the barrel and endcap EM calorimeters), to pass photon shower shape and electron rejection criteria, and to be isolated~\cite{Aad:2019tso,Aaboud:2018yqu}.


\section{Analysis strategy}
\label{sec:analysis_strategy}
 
In total, four sets of SRs are defined to target bottom squark pair-production or generic WIMP production in association with \bjets and are labelled SRX with X = A to D.
Each set of signal regions targets different values of \dmBN, the mass separation between the \sbottomone and \ninoone, or low and high dark matter mediator masses. The event selections defined for these regions all require the absence of baseline leptons, and exploit different techniques to improve the sensitivity to the target signal models. \SRA{} targets large values of \dmBN, and its definition resembles that used in Refs. \cite{SUSY-2011-15,Aad:2013ija,Aad:2015pfx,Aaboud:2016nwl,Aaboud:2017wqg}. \SRB, whose selection is mutually exclusive with that of \SRA{}, is designed to be optimal for $\splitSRBSRC\ \GeV < \dmBN < 200\ \GeV$, and uses a boosted decision tree  (BDT) \cite{Chen:2016:XST:2939672.2939785} as the final discriminant. \SRC targets signals with $\dmBN < \splitSRBSRC\ \GeV$, and exploits the information from the \softbvrt algorithm about the presence of vertices associated with low-\pt $b$-hadrons produced by the bottom squark decays. When deriving mass exclusion limits on bottom squarks or leptoquarks, \SRA{} and \SRB are statistically combined, and the analysis yielding the better of the expected \cls values  \cite{Read:2002hq} from the combined \SRA{}/\SRB and \SRC is used for each signal point. Finally, \SRD is optimised to target the dark matter models with scalar or pseudoscalar mediators by making use of a BDT.
 
For all signal regions, the SM background estimation is performed with a likelihood fit~\cite{Baak:2014wma} where the normalisation factors of the MC datasets corresponding to the SM processes  expected to contribute the most to the event yields in the SRs (\Zjets for all signal regions, \Wjets and \ttbar for \SRC) are left free to float. To aid their determination, dedicated control regions (CR) select events containing either one or two leptons, and having kinematic properties similar to events in the signal regions, but with negligible expected signal contributions. The quality of the background estimation is verified in dedicated validation regions (VR), designed to select events as similar as possible to those populating the SRs, while keeping signal contributions low. The likelihood is built as the product of Poissonian terms for each CR and, when assessing the discovery and exclusion sensitivity to new phenomena, SR bins. The effect of systematic uncertainties on the Poissonian expectation values is included through nuisance parameters assumed to have Gaussian probability distributions, as described in Section~\ref{sec:systematic_uncertainties}.
 
\subsection{Discriminating variables}
\label{sec:discriminating_variables}
 
Several kinematic variables built from the physics objects defined in the previous section are used to discriminate new physics from known SM background events. Variables which are used in many SRs are described here, while SR-specific variables are described in the corresponding SR sections below. Wherever necessary, final-state objects are labelled following a descending \pt ordering.
 
\begin{itemize}
\item \dphimin{n}: The minimum $\Delta\phi$ between any of the leading $n$ jets and \ptmiss. The background from multijet processes is characterised by small values of this variable.
\item \HTthree: It is defined as the scalar sum of the \pt of all jets excluding the leading two:
\[
\HTthree = \sum_{i \ge 3} (p_{\mathrm{T}}^{\mathrm{jet}})_{i}.
\]
The variable is used to reject events with extra-jet activity in signal regions targeting models characterised by small mass-splitting between the bottom squark and the neutralino.
\item \meff : It is defined as the scalar sum of the \pt\ of the jets and the \met, i.e.:
\[
\meff = \sum_i (p_{\mathrm{T}}^{\mathrm{jet}})_{i} + \met.
\]
The \meff\ observable is correlated with the mass of the directly pair-produced SUSY particles and
is employed as a discriminating variable, as well as in the computation of other composite observables.
\item \metsigObj : The global \met significance, calculated including parameterisations of the resolutions of all selected objects~\cite{ATLAS-CONF-2018-038}.  It is defined as follows:
 
\begin{equation*}
\metsigObj = \sqrt{\frac{|\ptmiss|^2}{\sigma_{\textup{L}}^2(1- \rho_{\textup{LT}}^2)}} .
\end{equation*}
 
Here $\sigma_{\textup{L}}$ is the total momentum resolution after being rotated into the longitudinal (parallel to the \ptmiss) plane.
The total momentum resolution of all jets and leptons, at a given \pt and $|\eta|$, is determined from parameterised Monte Carlo simulation in which the resolution measured in data is modelled well.
The quantity $\rho_{\textup{LT}}$ is a correlation factor between the longitudinal and transverse momentum resolution (again with respect to the \ptmiss) of each jet or lepton. The significance \metsigObj is used to discriminate between events where the \met arises from invisible particles in the final state and events where the \met arises from poorly measured particles (and jets).
 
\item $\mjj$ : The invariant mass of the two leading jets.
In events where at least one of the leading jets is \btagged, this variable helps to reduce the contamination from \ttbar\ events. It is referred to as $\mbb$ when the two leading \btagged jets are considered.
\item $\MT$ : The transverse mass of the lepton and the missing transverse momentum  is defined as:
\begin{equation*}
\MT = \sqrt{2 \ptlepsc \met - 2 \ptlepvec \cdot \ptmiss}
\end{equation*}
and is used in the CRs to suppress the contribution from fake and non-prompt leptons, which are normally characterised by low \MT values in multijet production events.
\item \mct: The contransverse mass~\cite{Tovey:2008ui} is the main discriminating variable in the \SRA{} signal regions. It is used to measure the masses of pair-produced heavy particles decaying semi-invisibly. For identical decays of two heavy particles (e.g.\ the bottom squarks decaying exclusively  as $\sbottomone \rightarrow b\nino$) into two visible particles $v_{1}$ and $v_{2}$ (the bottom quarks), and two invisible particles $X_1$ and $X_2$ (the $\nino$ for the signal), \mct\ is defined as
\begin{equation*}
\mct^{2}(v_1, \ v_2) = [\et(v_1)+\et(v_2)]^{2} - [\mathbf{p}_{\mathrm T}(v_1) - \mathbf{p}_{\mathrm T}(v_2)]^{2},
\end{equation*}
\noindent with $\et = \sqrt{\pt^{2}+m^{2}}$, and it has a kinematic endpoint at $\mct^{\mathrm{max}} = (m_{I}^2 - m_{X}^2)/m_{I}$, where $I$ is the initially pair-produced particle. This variable is extremely effective in suppressing the top quark pair production background ($I = t, X=W$), for which the endpoint is at 135~\GeV.
\item \minmtfour : This is the minimum of the transverse masses calculated using any of the leading four jets and the \ptmiss in the event.
For signal scenarios with low values of $\mct^{\mathrm{max}}$, this kinematic variable is an alternative discriminating
variable to reduce the \ttbar\ background.
\end{itemize}

\subsection{\SRA{} definition}
\label{sec:sbotA:SRs}
 
\SRA{} targets bottom squark pair production with large values of \dmBN. The selection criteria are summarised in Table~\ref{tab:SRA_summary}.
Only events with \met$>250$~\GeV\ are retained to ensure full efficiency of the online trigger selection and comply with the expected signal topology. To discriminate against multijet production, events where \ptmiss originates from the mismeasurement of a jet are suppressed with selections on $\dphimin{4}$ and  $\met/\meff$.  The final state is expected to contain two \bjets from the two bottom squark decays. A veto on large hadronic activity (implemented by rejecting events with a fourth jet of significant \pt) is imposed to suppress mostly events from SM \ttbar production. SM $\Wjets$ and $\Zjets$ production, where $\bjets$ are produced mainly via gluon splitting, is suppressed by a selection on \mbb. Finally, selections on \meff and \mct are applied to maximise the sensitivity to the signal. When excluding specific models of bottom squark production, a two-dimensional binning in \mct and \meff is applied. Five mutually exclusive regions ($\mct\in [250, 350),\;[350,450),\;[450,550),\;[550,650)$ and $[650,\infty)$, with all units in \GeV) denoted by \SRA{mctX}, where X is the bin lower bound, are used. \SRA{mct250} is subdivided into five bins of \meff, starting from $\meff>500\ \GeV$ and increasing in steps of 200 \GeV, with the last bin including all events with $\meff > 1300\ \GeV$. \SRA{mct350} and \SRA{mct450} are both defined with two bins of \meff ($[0.5\ \TeV,1\ \TeV),\;[1\ \TeV,\infty)$ and $[1\ \TeV,1.5\ \TeV),\;[1.5\ \TeV,\infty)$ respectively). Due to the relatively small number of events selected by the highest two \mct bins, a single selection $\meff > 1.0\,(1.5)\ \TeV$ is applied in \SRA{mct550} (\SRA{mct650}) respectively. When assessing the model-independent discovery significance against the background-only hypothesis (see Section~\ref{sec:results_and_interpretation}), five discovery regions, named \SRA{mctXi} are defined by removing any binning in \meff.
 
\begin{table}[!htpb]
\caption{\SRA{} signal, control and validation region definitions. Pink cells  for the control and validation regions' columns indicate which selections ensure that the regions are orthogonal to the SR.}
\vspace{0.5cm}
\label{tab:SRA_summary}
\centering
\resizebox{\columnwidth}{!}{
\begin{tabular}{l | l | c|c|c|c|c|c}
\toprule
Variable          &                    & \SRA          & \CRZA  &  \VRAA  & \VRAB  & \VRAC    & \VRAD   \\
\midrule
Number of baseline leptons &     & 0              & \cellcolor{red!25} 2   & \multicolumn{4}{c}{0}  \\
Number of high-purity leptons &     & --              & 2 SFOS    & \multicolumn{4}{c}{--}\\
$\pt (\ell_1)$ & [\GeV]                   & --             & $> 27$    & \multicolumn{4}{c}{--}            \\
$\pt (\ell_2)$ & [\GeV]                   & --             & $> 20$    & \multicolumn{4}{c}{--}           \\
\MT    & [\GeV]                          & --             & $>20$     & \multicolumn{4}{c}{--}  \\
\mll & [\GeV]                            & --             & $[81, 101]$  & \multicolumn{4}{c}{--}            \\
\midrule
Number of jets &                        & \multicolumn{6}{c}{$\in[2,4]$}         \\
Number of $b$-tagged jets &               &   \multicolumn{6}{c}{2}      \\
$j_1$ and $j_2$ $b$-tagged &           & \multicolumn{6}{c}{\ding{51}}     \\
$\pt (j_1)$ & [\GeV]                   & \multicolumn{6}{c}{$>$ 150}~ \\
$\pt (j_2)$ & [\GeV]                   & \multicolumn{6}{c}{$>$ 50} \\
$\pt (j_4)$ & [\GeV]                   & \multicolumn{6}{c}{$<$ 50} \\
\dphimin{4}  & [rad]                        &  \multicolumn{6}{c}{$>$ 0.4}    \\
\midrule
\met & [\GeV]                            & $>$ 250        & $<100$  & \multicolumn{4}{c}{$>$ 250}\\
\metprime & [\GeV]                      & --             & $>250$ &  \multicolumn{4}{c}{ --}\\
$\met / \meff$ &                        & $>$ 0.25       & -- &\multicolumn{4}{c}{--}\\
$\metprime / \meff$ &                    & --             & $>0.25$ & \multicolumn{4}{c}{--}\\
\midrule
\mbb & [\GeV]                            & \multicolumn{2}{c|}{$> 200$} & \cellcolor{red!25} $< 200$ & $> 200$ & \cellcolor{red!25}$< 200$ & $> 200$ \\
\mct & [\GeV]                            & \multicolumn{2}{c|}{$> 250$} & $> 250$ & \cellcolor{red!25}$[150, 250]$ & $> 250$ & \cellcolor{red!25}$[150, 250]$\\
\meff & [\GeV]                           & \multicolumn{2}{c|}{$> 500$} & \multicolumn{2}{c|}{$[500, 1500]$} & \multicolumn{2}{c}{$>1500$} \\
\bottomrule
\end{tabular}
}
\end{table}

\subsection{\SRB definition}
\label{sec:sbotB:SRs}
 
If $\dmBN<200$~\GeV, selections based on the \mct\ and \mbb\ variables are no longer effective and a multivariate approach is preferred to separate the signal from SM production processes. A BDT is implemented by making use of the \verb|XGBoost| (XGB) framework \cite{Chen:2016:XST:2939672.2939785}. The training procedure used events that pass the selection specified in Table~\ref{tab:SRB_summary} (with the exception of the BDT output score) and are classified in four different categories: three corresponding to the main backgrounds processes (\ttbar, \Zjets, \Wjets production), and one grouping together semi-compressed signal samples ($\dmBN\leq 200\ \GeV$, where the event selection suppresses the acceptance for samples with $\dmBN\leq 30\ \GeV$ ), for scalar bottom squark masses $m_{\sbottomone} < 800\ \GeV$. A \textit{one vs rest} multi-classification procedure was used: for each classifier, the class is fitted against all the other classes producing output scores containing the predicted probability of an event being in each class. The output score \XGB denotes the signal classifier output score and is used in the definition of the signal region. The rotational invariance of event topologies in the transverse plane is exploited by rotating the azimuthal angles of all final-state objects so that \met has $\phi(\ptmiss)=0$.  The variables used in the training are the momentum vectors of the jets, the \btagging information, and other event-level variables (\meff, \metsigObj, \mct, \minmtfour and \dRbb). The highest-ranked variables after training are \minmtfour and the transverse momenta of the first three jets in the event.
 
The full selection of \SRB is defined in Table~\ref{tab:SRB_summary}. An upper bound on \mct ensures that the selection is orthogonal to \SRA{}. When assessing the exclusion sensitivity for the signal-plus-background hypothesis for specific BSM models, four \XGB bins are used in the likelihood fit ($[0.75,0.80),\;[0.80,0.85),\;[0.85, 0.90),\;[0.90,1]$).
 
\begin{table}[!htpb]
\caption{\SRB signal, control and validation region definitions. Pink cells  for the control and validation regions' columns indicate which selections ensure that the regions are orthogonal to the SR.}
\vspace{0.5cm}
\label{tab:SRB_summary}
\centering
\begin{tabular}{l | l |c|c|c}
\toprule
Variable       &                       & \SRB     &  \CRZB    & \VRB \\
\midrule
Number of baseline leptons &     & 0              & \multicolumn{2}{c}{\cellcolor{red!25} 2}  \\
Number of high-purity leptons &     & --             & \multicolumn{2}{c}{2 SFOS } \\
$\pt (\ell_1)$ & [\GeV]                   & --             & \multicolumn{2}{c}{$> 27$     } \\
$\pt (\ell_2)$ & [\GeV]                   & --             & \multicolumn{2}{c}{$> 20$    } \\
\mll & [\GeV]                           & --             & \multicolumn{2}{c}{$[76, 106]$ } \\
\MT  & [\GeV]                          & --             & \multicolumn{2}{c}{$>20$} \\
\midrule
Number of jets &                        & \multicolumn{3}{c}{$\in[2,4]$}         \\
 
Number of $b$-tagged jets &               &   \multicolumn{3}{c}{2}      \\
 
$\pt (j_1)$ & [\GeV]                   & \multicolumn{3}{c}{$>$ 100} \\
$\pt (j_2)$ & [\GeV]                   & \multicolumn{3}{c}{~~$>$ 50} \\
\dphimin{4}  & [rad]                        &  \multicolumn{3}{c}{~$>$ 0.4}    \\
$j_1$ not \btagged             &            &    -- &  \ding{51} & --   \\
\midrule
\met & [\GeV]                            & $>$ 250        & \multicolumn{2}{c}{$<$ 100}\\
\metprime & [\GeV]                      & --             & \multicolumn{2}{c}{$>$ 250} \\
\midrule
\mct &  [\GeV] &                      \multicolumn{3}{c}{$<250$}          \\
\midrule
\XGB  &       &                      $>0.85$ & $[0.3,0.63]$ &  $>0.63$ \\
\bottomrule
\end{tabular}
\end{table}

\subsection{\SRC definition}
\label{sec:sbotC:SRs}
 
\SRC targets events where a bottom squark pair is produced recoiling against a high-\pt initial-state-radiation (ISR) jet and $\dmBN < \splitSRBSRC\ \GeV$. In the boosted bottom squark decay, the boost is mostly transferred to \ninoone because of its mass. It is because of such boost that the \met satisfies the trigger requirements, while the bottom quarks are instead expected to have low \pt. Three mutually exclusive signal regions, based on the number of \btagged jets and \softbvrt-identified vertices (\nvtx), are defined: \SRCA, two \bjets; \SRCB, one \bjet and at least one \softbvrt vertex; and \SRCC, no \bjets and at least one \softbvrt vertex. The three regions offer complementary sensitivity depending on \dmBN, and are statistically combined when stating the sensitivity for exclusion of bottom squark pair production models. They all exploit the topological and kinematic features of the signal by requiring large \met and a high-\pt, non-\btagged leading jet, and vetoing on additional hadronic activity by imposing an upper bound on \HTthree.  The following variables are used to better extract the signal from the SM background:
 
\begin{itemize}
\item The bottom quarks coming from the bottom squark decay are expected to be produced centrally in pseudorapidity, angularly close to each other and nearly back-to-back to the ISR jet. This is exploited in \SRCB and \SRCC with selections on the angular separation in the transverse plane between the leading jet and the \bjet or \softbvrt vertex, and on the pseudorapidity of the \softbvrt vertex,  \etavtx.
 
\item The \pt of the leading ISR jet is expected to be significantly higher than that of the second jet, expected to come from the bottom squark decay. Therefore the variable
\begin{equation*}
{\cal A} = \frac{\pt(j_1) - \pt(j_2)}{\pt(j_1) + \pt(j_2)}
\end{equation*}
\noindent is expected to take values close to one for the signal, while it is expected to have a wider distribution for the background. This variable is not used in \SRCC, where a jet coming from the bottom squark decay cannot be identified.
\item The vertex mass (\mvtx) and \pt (\ptvtx) are useful in rejecting events where the vertex is due to a $c$-hadron decay or to a random track crossing. For these {\em fake} vertices the values of both variables tend to be lower than for vertices originating from $b$-hadron decays.
\end{itemize}
 
The full list of selections applied to these variables and to other variables introduced in Section~\ref{sec:discriminating_variables} is shown in Table~\ref{tab:SRC_SR_summary}.
To further enhance the exclusion sensitivity, two different bins in \met are defined ($\met \in[500\ \GeV, 650\ \GeV),\;[650\ \GeV, \infty)$ for \SRCA and $\met \in[400\ \GeV, 600\ \GeV),\;[600\ \GeV, \infty)$ for \SRCB and \SRCC).
 
\begin{table}[!htpb]
\caption{\SRC signal and validation region definitions. Pink cells  for the validation regions' columns indicate which selections ensure that they are orthogonal to the corresponding SR.
}
\vspace{0.5cm}
\label{tab:SRC_SR_summary}
\centering
\resizebox{\columnwidth}{!}{
\begin{tabular}{l|l|c|c|c|c|c|c}
\toprule
 
Variable                                         &           & \SRCA                                     & \SRCB     & \SRCC        &  \VRCA    & \VRCB  & \VRCC    \\
\midrule
Number of jets                                &           & \multicolumn{6}{c}{$\in[2,5]$}         \\
$j_1$ not \btagged                         &           & \multicolumn{6}{c}{\ding{51}}     \\
Number of baseline leptons  &           &  \multicolumn{6}{c}{0}  \\
Number of \btagged jets                 &           &   $\ge 2$                                  & 1            & 0                   & $\ge 2$     & 1            & 0      \\
\midrule
 
\nvtx                                                &           & $\ge 0$                                    & $\geq 1$ & $\ge 1$        & $\ge 0$     & $\geq 1$ & $\ge 1$ \\
\mvtx 				                   & [\GeV] & $-$                                           & ~~~$> 0.6$   &  ~~~$> 1.5$       & $-$            & ~~~$> 0.6$   &  ~~~$> 1.5$ \\
\ptvtx 				                   & [\GeV] & $-$                                           & $> 3$      & $>5$            & $-$            & $> 3$      & $>5$ \\
\midrule
 
$\pt (j_1)$                                         & [\GeV] & $> 500$                                   & $> 400$  & $> 400$       & \cellcolor{red!25} $< 500$     & $> 400$ & $> 400$ \\
\met                                                 & [\GeV] & $> 500$                                   & $> 400$  & $> 400$       & \cellcolor{red!25} $<500$      & $> 400$  & $> 400$ \\
\HTthree                                          & [\GeV] & $-$                                           & ~~$< 80$    & ~~$< 80$         & $-$             & ~~$< 80$    & ~~$< 80$ \\
$\cal{A}$                                          &           & $> 0.80$~                                  & $> 0.86$~ & $-$               & $[0.8,0.9]$ & $> 0.86$~ & - \\
\mjj                                                  & [\GeV] & $> 250$                                    & $>250$    & $-$              & \cellcolor{red!25}$[150,250]$ & $> 250$ & - \\
\dphiISRb                                        & [rad]  & $-$                                          & $ > 2.2$   & $-$               & $-$             & \cellcolor{red!25} $< 2.2$   & $-$\\
\dphiISRvtx                                     &  [rad]  & $-$                                            & $ - $         & $> 2.2$       & $-$             & $ - $        & \cellcolor{red!25} $< 2.2$\\
$|\etavtx|$                                       &           & $-$                                           & $< 1.2$    & $ < 1.2$     & $-$             & \cellcolor{red!25} $ > 1.2$  & \cellcolor{red!25} $ > 1.2$ \\
\bottomrule
\end{tabular}
}
\end{table}
 
\subsection{\SRD definition}
\label{sec:DM:SRs}
 
Two signal regions target low- and high-mediator-mass dark matter signals, and are named \SRDA and \SRDB, respectively:
\SRDA is optimised for mediator masses from $10$ to $100$~\GeV, while \SRDB is optimised for mediator masses from $200$ to $500$~\GeV.
A common preselection is applied including the requirement of two \bjets in the final state.
The thresholds for the missing transverse momentum and the \pt of the leading jet are kept as low as possible via
a two-dimensional requirement selecting events on the trigger plateau,
i.e.\ $(\pt (j_1) - 20\;\GeV)(\met - 160\;\GeV) > 5000\;\GeV^2$.
Then BDTs are trained to discriminate between the three most relevant background processes
(top pair production, \Wjets, \Zjets) and two sets of kinematically similar signal models
which are characterised by either low or high mediator mass.
This results in six BDT discriminants, denoted by $w^{\mathrm{X}}_\mathrm{Y}$,
where $\mathrm{X}$ and $\mathrm{Y}$ are the background process and signal mass range used in the training, respectively.
The BDT discriminants have ranges of $[-1, 1]$ with the more positive values being more signal-like.
In addition to some of the variables listed in Section~\ref{sec:discriminating_variables},
the following variables are used specifically in \SRD:
 
\begin{itemize}
\item \HT: the scalar sum of the jet transverse momenta. The ratio of the leading jet \pt to \HT is used in the signal region selection.
\item $\delta^+$, $\delta^-$: angular variables that exploit the topology of the event~\cite{SUSY-2016-18}.
They are defined as two linear combinations of \dphimin{3} and the azimuthal separation between the \bjets, $\Delta\phi_{bb}$.
\begin{align*}
\delta^- &= \dphimin{3} - \Delta\phi_{bb}, \\
\delta^+ &= |\dphimin{3} + \Delta\phi_{bb} - \pi|.
\end{align*}
\end{itemize}
 
These variables are used in the training of the different BDTs together with the \pt of the
leading $b$-jet and of the second and third jets in the event, \met,  \metsigObj, \dphimin{3}, and
\mct computed using the two leading jets. The most discriminating variables are \dphimin{3} and the ratio of the leading jet \pt to \HT.
The signal region selections are detailed in Table~\ref{tab:SRD_summary}.
A final discriminating variable $\cosbb$~\cite{Haisch:2016gry} is considered: it is defined as
\begin{equation*}
\cosbb =\left| \tanh \frac{\Delta \eta \left( b_1, b_2 \right)}{2}\right|.
\end{equation*}
When excluding models of DM production, the SRDs are further divided  into five equal bins of width $0.2$ in the $[0,1]$ range of \cosbb. When assessing the model-independent discovery significance against the background-only hypothesis, a single bin in \cosbb defined by  $\cosbb > 0.6\,(0.8)$ is used in \SRDA (\SRDB).
 
\begin{table}[!htpb]
\caption{\SRD signal, control and validation region definitions. Pink cells  for the control and validation regions' columns indicate which selections ensure that they are orthogonal to the corresponding SR.
}
\vspace{0.5cm}
\label{tab:SRD_summary}
\centering
\resizebox{\columnwidth}{!}{
\begin{tabular}{l|l|c|c|c|c|c|c}
\toprule
Variable      &                     & \SRDA & \SRDB                &  \CRZDA & \CRZDB                     & \VRZDA & \VRZDB \\
 
\midrule
 
Trigger plateau  &            &  \multicolumn{6}{c}{$(\pt (j_1) - 20\;\GeV)(\met - 160\;\GeV) > 5000\;\GeV^2$}             \\
\njets           &            &  \multicolumn{6}{c}{2--3}             \\
\nbjets          &            &  \multicolumn{6}{c}{$\ge 2$} \\
$\pt(j_1)$ & [\GeV]                 &  \multicolumn{6}{c}{$>100$} \\
$\pt(j_2)$ & [\GeV]                &  \multicolumn{6}{c}{~~$>50$} \\
$\dphimin{3}$    & [rad]           &  \multicolumn{6}{c}{$> 0.4$}            \\
\metsigObj     &                       &  \multicolumn{6}{c}{$>7$}~~               \\
$\pt(j_1)/\HT$          &                  &  \multicolumn{6}{c}{$>0.7$}             \\
 
\midrule
Number of baseline leptons &  & \multicolumn{2}{c|}{0}              &  \multicolumn{2}{c|}{\cellcolor{red!25} 2}            & \multicolumn{2}{c}{0}           \\
Number of high-purity leptons &  & \multicolumn{2}{c|}{--}              & \multicolumn{2}{c|}{2 SFOS}            & \multicolumn{2}{c}{--}           \\
$\pt(\ell_1)$ &  [\GeV]            & \multicolumn{2}{c|}{--}              & \multicolumn{2}{c|}{$> 27$}       & \multicolumn{2}{c}{--}\\
$\pt(\ell_2)$ & [\GeV]            & \multicolumn{2}{c|}{--}              & \multicolumn{2}{c|}{$> 20 $}   & \multicolumn{2}{c}{--}            \\
\MT&  [\GeV]                           & \multicolumn{2}{c|}{--}              & \multicolumn{2}{c|}{$> 20 $}   & \multicolumn{2}{c}{--}               \\
$\mll$ & [\GeV]                          & \multicolumn{2}{c|}{--}             & \multicolumn{2}{c|}{$ [81,101]$}         & \multicolumn{2}{c}{--} \\
\midrule
\metprime & [\GeV] & \multicolumn{2}{c|}{--} & \multicolumn{2}{c|}{$> 180$} & \multicolumn{2}{c}{--} \\
\met  & [\GeV]                          & \multicolumn{2}{c|}{$> 180$}              & \multicolumn{2}{c|}{$ < 100$}     & \multicolumn{2}{c}{$ > 180 $}         \\
\midrule
$\BDTttDA$ & & $>0$ & --             & \multicolumn{2}{c|}{--}                    & $>0$          & --   \\
$\BDTZDA$  &  & $>0$ & --              & $>0$ & --                    & {\cellcolor{red!25}$[-0.2,0]$}  & --\\
$\BDTWDA$ &  & $>0$ & --              & \multicolumn{2}{c|}{--}                    & $>0$          & --\\
$\BDTttDB$ &  & -- & $>0$            & \multicolumn{2}{c|}{--}                     & --             & $>0$ \\
$\BDTZDB$ &  & -- & $>-0.1$         & -- & $>-0.1$                    & --     & {\cellcolor{red!25}$[-0.3,-0.1]$}\\
$\BDTWDB$ &  & -- & ~~$>-0.05$        & \multicolumn{2}{c|}{--}                    & --     & $>-0.05$\\
\bottomrule
\end{tabular}

}
\end{table}

\subsection{Control and validation region definition}
 
Event selections kinematically similar to those of the signal regions are defined for the control regions, which are characterised by negligible expected signal contributions for the BSM models considered. Contrary to the SRs, such CRs rely on the presence of either one or two same-flavour opposite-sign (SFOS) high-purity electrons or muons (generically denoted by $\ell$), and are defined such that their event yield is dominated by one specific SM production process. They are part of the likelihood fit, where they are key to determining the value of the free-floating normalisation parameter associated with the MC prediction of the dominant background process.
 
The SM background yield is dominated in most signal regions by \Zjets production followed by $Z\rightarrow \nu\bar{\nu}$. For each signal region, a corresponding control region (CRz) with two SFOS leptons is defined, with an invariant mass of the lepton pair close the $Z$ boson mass: the kinematic properties of the events populating such a control region are expected to be very similar to those of events in the signal region. The full definition of the control region selection needs to take into account the lower branching ratio of $Z\rightarrow \ell \ell$ relative to $Z\rightarrow \nu \bar{\nu}$: the selection is therefore close, but not identical, to that of the signal region. After having rejected events with high \met values to suppress contributions from dileptonic \ttbar production, the \pt of the leptons is added vectorially to the \ptmiss to mimic the expected missing transverse momentum spectrum of $Z \rightarrow \nu \bar{\nu}$ events, and is denoted in the following by \metprime. All variables constructed from \met are recomputed using \metprime instead, including the BDT scores used in regions B and D. The selections corresponding to the control regions associated with \SRA{} and \SRB, named \CRZA and \CRZB, are shown in Tables~\ref{tab:SRA_summary} and \ref{tab:SRB_summary}, respectively. Those corresponding to the control regions associated with \SRDA and \SRDB, named \CRZDA and \CRZDB, are shown in Table~\ref{tab:SRD_summary}. In the case of \SRC, one \Zjets control region is defined for each of \SRCA, \SRCB and \SRCC: they are named \CRZCA, \CRZCB and \CRZCC respectively, and their selection is shown in Table~\ref{tab:SRC_CR}.
 
\begin{table}[!htpb]
\caption{\SRC control region definitions. Pink cells  for the control regions' columns indicate which selections ensure that they are orthogonal to the corresponding SR. 
}
\vspace{0.5cm}
\label{tab:SRC_CR}
\centering
\resizebox{\columnwidth}{!}{
\begin{tabular}{l|c|c|c|c|c|c|c}
\toprule
Variable &          & \CRTC& \CRWCB & \CRWCC & \CRZCA & \CRZCB & \CRZCC \\
\midrule
$j_1$ not \btagged   &  & \multicolumn{6}{c}{\ding{51}}  \\
Number of high-purity leptons  &  & \multicolumn{3}{c|}{\cellcolor{red!25} $1$} & \multicolumn{3}{c}{\cellcolor{red!25} $2$ SFOS}\\
\HTthree & [\GeV]   & \multicolumn{6}{c}{$<80$}\\
\midrule
$\pt(j_1)$ &[\GeV]      & \multicolumn{3}{c|}{$>400$} & $>300$ &\multicolumn{2}{c}{$>400$} \\
\MT    & [\GeV]        & \multicolumn{3}{c|}{$[20,120]$} & \multicolumn{3}{c}{$-$} \\
\mll  & [\GeV]      &  \multicolumn{3}{c|}{$-$} & \multicolumn{3}{c}{$[81, 101]$ }\\
\met & [\GeV]  & \multicolumn{3}{c|}{$> 400$} & \multicolumn{3}{c}{$<100$} \\
\metprime & [\GeV] & \multicolumn{3}{c|}{$-$} & $>250$ & \multicolumn{2}{c}{$>400$} \\
\midrule
$\cal{A}$        &          & $> 0.5$~   & $> 0.8$~  & $-$ & $> 0.5$~   & $> 0.8$~  & $-$\\
\mjj    & [\GeV]       & $>250$    & $>250$   & $-$ & $-$ & $>250$ & $-$ \\
\nbjets    &         & $\ge 2$~~~~   & $1$    & $0$ & $\ge 2$~~~~   & $1$    & $0$\\
\midrule
\nvtx       &         & $-$       & $\ge 1$~~~  & $\ge 1$~~~ & $-$       & $\ge 1$~~~  & $\ge 1$~~~ \\
\mvtx   & [\GeV]         & $-$       & $>0.6$   & $>1.5$ &  $-$       & $>0.6$   & $>1.5$\\
\ptvtx  & [\GeV]        & $-$       & $> 3$~~~    & $>5$~~~ & $-$       & $> 3$~~~    & $>5$~~~ \\
\bottomrule
\end{tabular}

}
\end{table}

The production of \Wjets and, to a lesser extent, top quarks, also results in important backgrounds in SRC. A set of control regions (CRt and CRw) is defined, all containing exactly one high-purity lepton in the final state. The zero-lepton signals considered for the signal region optimisation do not contaminate the one-lepton control regions.
However, potential signal contributions from possible related BSM signal production (e.g.\ top squark pairs) or from third-generation leptoquarks are rejected by imposing an upper bound on the transverse mass of the lepton and the missing transverse momentum, \MT.
 
A common top control region containing two \btagged jets and no \softbvrt vertex, named \CRTC, and two \Wjets control regions containing at least one \softbvrt vertex and, respectively, one (\CRWCB) and no (\CRWCC) \btagged jets are defined and summarised in Table~\ref{tab:SRC_CR}. The definition of a \Wjets control region containing two \btagged jets was considered, but it was found too difficult to obtain a satisfactory \Wjets purity because of contamination from top quark production.

Finally, a series of validation regions is defined, with the purpose of evaluating the quality of the background estimation after the likelihood fit. They are characterised by an expected signal contamination below 10\%, and they are obtained by inverting one or more signal region variable selections. They are defined in Tables~\ref{tab:SRA_summary}, \ref{tab:SRB_summary}, \ref{tab:SRC_SR_summary} and  \ref{tab:SRD_summary}
 
\FloatBarrier

\FloatBarrier

\section{Systematic uncertainties}
\label{sec:systematic_uncertainties}
 
The effects of several sources of systematic uncertainty on the signal and background estimates are introduced in the likelihood fit through nuisance parameters that affect the expectation values of the Poissonian terms for each CR and SR  bin. Each nuisance parameter's probability density function is described by a Gaussian distribution whose standard deviation corresponds to a specific experimental or theoretical modelling uncertainty. The preferred value of each nuisance parameter is determined as part of the likelihood fit. The fits performed do not significantly alter or constrain the nuisance parameter values relative to the fit input.
 
Jet energy scale and resolution uncertainties are derived as a function of the jet $\pt$ and $\eta$, jet flavour, and pile-up conditions, using a combination of data and simulated events through measurements of jet response asymmetry for several processes, as detailed in Refs.~\cite{PERF-2016-04, PERF-2011-04}. The impact of uncertainties on the efficiencies and mis-tag rates of the \btagging algorithm is estimated by varying, as a function of $\pt$, $\eta$ and jet flavour, the scale factors used to correct the MC simulation, within a range reflecting the uncertainty in their measurement~\cite{Aad:2019aic}. Similarly, the impact of the uncertainty on the MC modelling of the efficiency and fake rate for the \softbvrt vertex reconstruction is estimated by varying the corresponding scale factors within the uncertainty associated with their determination (about 6\% for the efficiency and 30\% for the fake rate).
Uncertainties connected with the lepton reconstruction and identification are included in the fit, and they are found to have a negligible impact. All uncertainties in the final-state object reconstruction are propagated to the reconstruction of the \met, including an additional one  taking into account uncertainties in the scale and resolution of the soft term.
 
Uncertainties in the modelling of the SM background processes from MC simulation are taken into account. They are assumed to be fully correlated across signal regions, but uncorrelated between different processes. An alternative correlation model, where the uncertainties are assumed to be uncorrelated across signal regions, leads to a small increase in the final yield uncertainty, but to no significant change in the mass and cross-section limits obtained.
 
Several contributions to the uncertainty in the theoretical modelling of \ttbar and single top production are considered. The uncertainty due the choice of hard-scattering generator and matching scheme is evaluated by comparing the nominal sample with a sample generated with \MGMCatNLO and a shower starting scale $\mu_{\mathrm q} = \HTgen/2$. The uncertainty due to the choice of parton shower and hadronisation model  is evaluated by a comparison with a sample generated with \POWHEGBOX interfaced to \HERWIG 7~\cite{Bahr:2008pv,Bellm:2015jjp}, using the H7UE set of tuned parameters~\cite{Bellm:2015jjp}. Variations of the renormalisation and factorisation scales, the initial- and final-state radiation parameters and PDF sets are also considered~\cite{ATL-PHYS-PUB-2018-009}. Uncertainties on the interference between the single top $Wt$ and $\ttbar$ production have negligible impact on the analysis results and are not included.
 
Uncertainties in the modelling of \Zjets\  and \Wjets~\cite{Bothmann:2016nao} are evaluated by using 7-point variations of the renormalisation and factorisation scales by factors of 0.5 and 2. The matching scale between the matrix element and parton shower calculation, and the resummation scale for soft gluon emission, are also varied by factors of 0.5 and 2. As no Monte Carlo generator has been found to accurately describe $Z+b\bar{b}$ production in all observables~\cite{Aad:2020gfi}, nor are these discrepancies accounted for by scale variations, an uncertainty due to the choice of generator is evaluated by comparing the nominal samples with those produced using \AMCatNLOV{2.3.3}+~\PYTHIA. After constraints from the control regions these variations are found to be relevant only in \SRD, where modelling uncertainties dominate the systematic effect on the shape of the \cosbb distribution.
 
The impact of the most relevant background systematic uncertainties in the different signal regions is shown in Figure~\ref{fig:systUnc}. Modelling uncertainties of the $\Zjets$ process dominate the signal regions' uncertainties, while the most important experimental uncertainties are those related to the jet energy scale.

\begin{figure}[!h]
\begin{center}
\includegraphics[width=0.6\textwidth, angle=90]{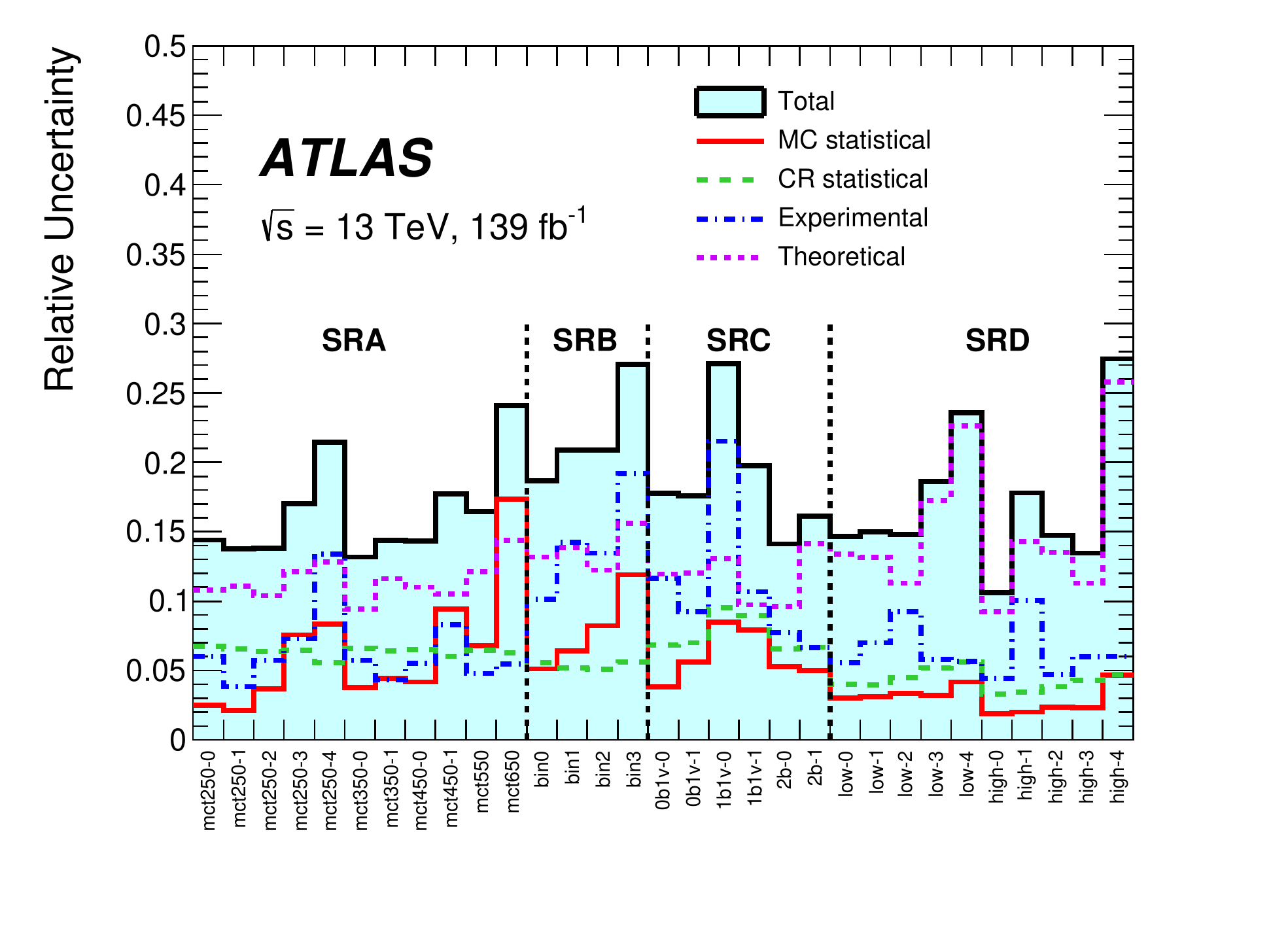}
\end{center}
\caption{Summary of the post-fit relative systematic uncertainties of the various signal region yields, also split by component.}
\label{fig:systUnc}
\end{figure}

 
\section{Results and interpretation}
\label{sec:results_and_interpretation}
 
Different likelihood fits are run when assessing the accuracy of the SM background determination ({\em background-only} fit), when computing the $p$-value of the SM-only hypothesis ({\em model-independent} fit) and when evaluating the confidence level for excluding a specific BSM hypothesis ({\em model-dependent} fit)~\cite{Baak:2014wma}.
 
In the background-only fit, only the control regions are used in the likelihood, and the predicted post-fit level of background is compared with the observed yields in the corresponding VRs and SRs. Three distinct fits are run for the combination of \SRA{} and \SRB, for \SRC and for \SRD. In the \SRA{}/\SRB and \SRD fits, only the normalisation of the \Zjets MC background prediction is left free to float. For \SRC, a combined fit is run including \SRCA, \SRCB and \SRCC: one common normalisation factor is applied to the \ttbar and single-top contributions; one normalisation factor is applied to the \Wjets MC predictions in all regions with one or more \btagged jets,
while an independent one is applied to those with no \btagged jets;
and, finally, three independent normalisation factors are applied to the \Zjets MC predictions depending on the number of \btagged jets.
\begin{figure}[!h]
\begin{center}
\includegraphics[width=0.7\textwidth]{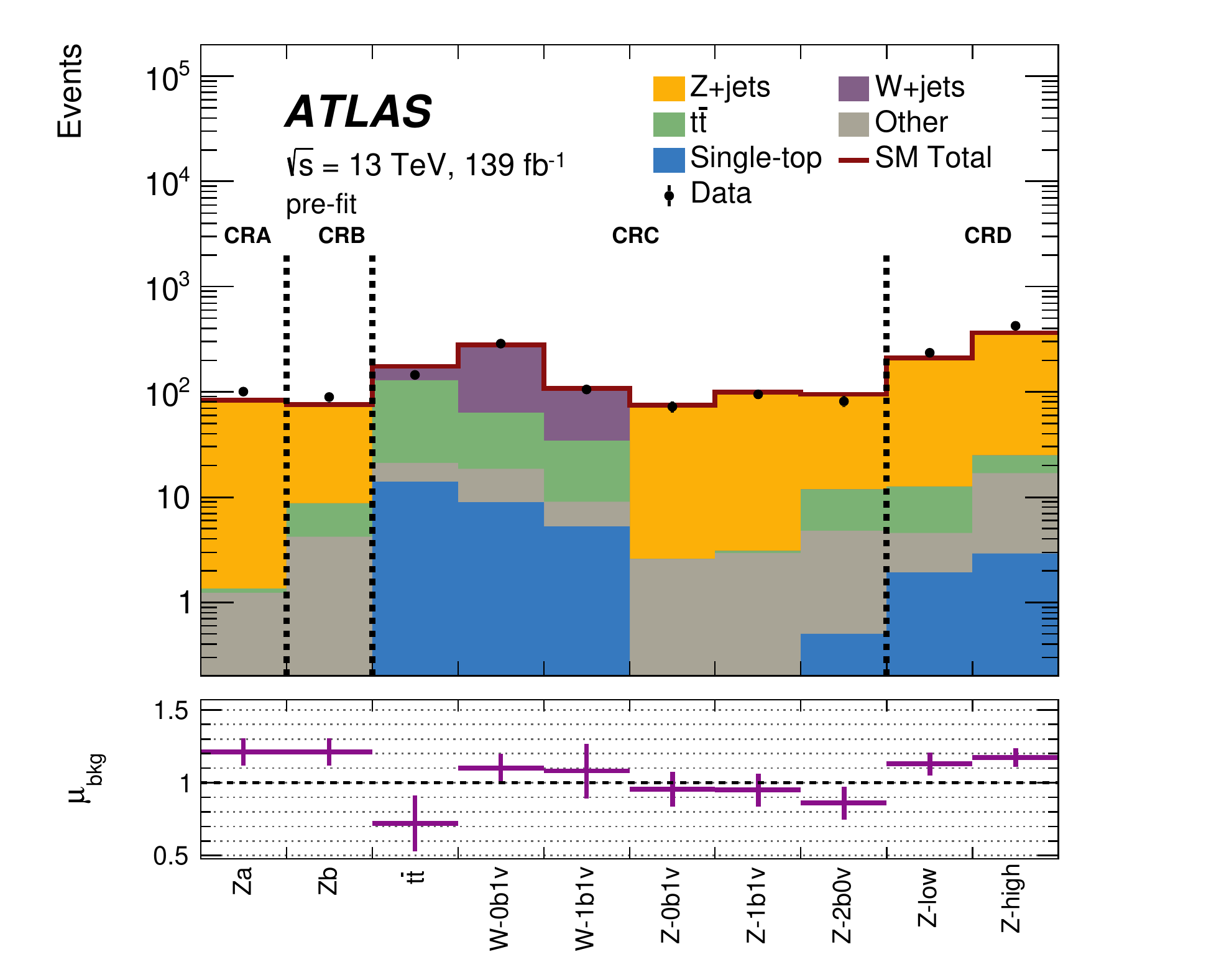}
\end{center}
\caption{Comparison between observed and pre-fit predicted background yields for the control regions. In the ratio plot, $\mu_\mathrm{bkg}$ represents the value (with uncertainty) of the free-floating normalisation parameter for the dominant background process in each region obtained with the background-only fit. The ``Other'' category includes contributions from diboson and $\ttbar+W/Z/H$ production.}
\label{fig:pull_CR}
\end{figure}
 
\begin{figure}[!h]
\begin{center}
\includegraphics[width=0.7\textwidth]{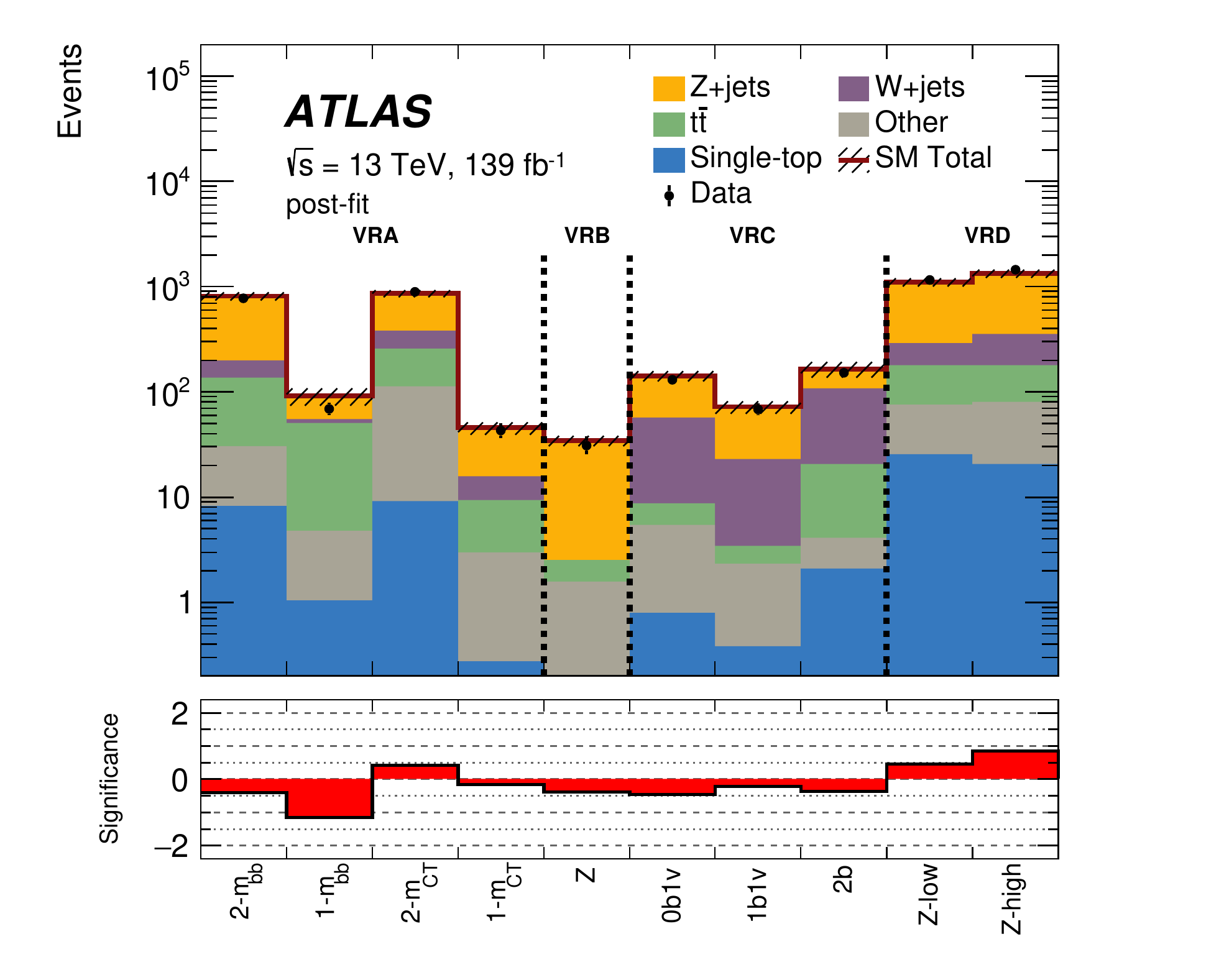}
\end{center}
\caption{Comparison between observed and post-fit predicted background yields for the validation regions. The ratio plot represents the statistical significance \cite{Cousins:2007bmb} of the discrepancy between the observed and predicted value. The ``Other'' category includes contributions from diboson and $\ttbar+W/Z/H$ production. The shaded band represents the total uncertainty of the background prediction. }
\label{fig:pull_VR}
\end{figure}
 
The background-only pre-fit event yields are compared with the observation for all CRs  in Figure~\ref{fig:pull_CR}, while Figure~\ref{fig:pull_VR} shows the background-only post-fit prediction compared with the observation for the VRs. Agreement is observed in the validation regions. Distributions of relevant variables in the validation regions were inspected in detail,
and were found to be consistent with the post-fit predictions. The corresponding SR yield comparison is shown in Figure~\ref{fig:pull_SR}. The largest background contribution in all SRs arises from \Zjets~ production,  while \Wjets and top quark production contributions are minor, except in \SRC. Other background sources are \ttbar+$W/Z$, \ttbar+$H$, and diboson production. Multijet production was found to be negligible in every region considered. No significant deviation from the background-only prediction is observed in any of the signal regions. An excess of about one standard deviation, covered by the largely correlated systematic uncertainty, is observed in the \SRDA bins. Figures~\ref{fig:SR_N1s_1} and~\ref{fig:SR_N1s_2} show comparisons between the observed data and the post-fit background predictions for relevant kinematic distributions for the SRs.
 
\begin{figure}[!h]
\begin{center}
\includegraphics[width=0.6\textwidth, angle=90]{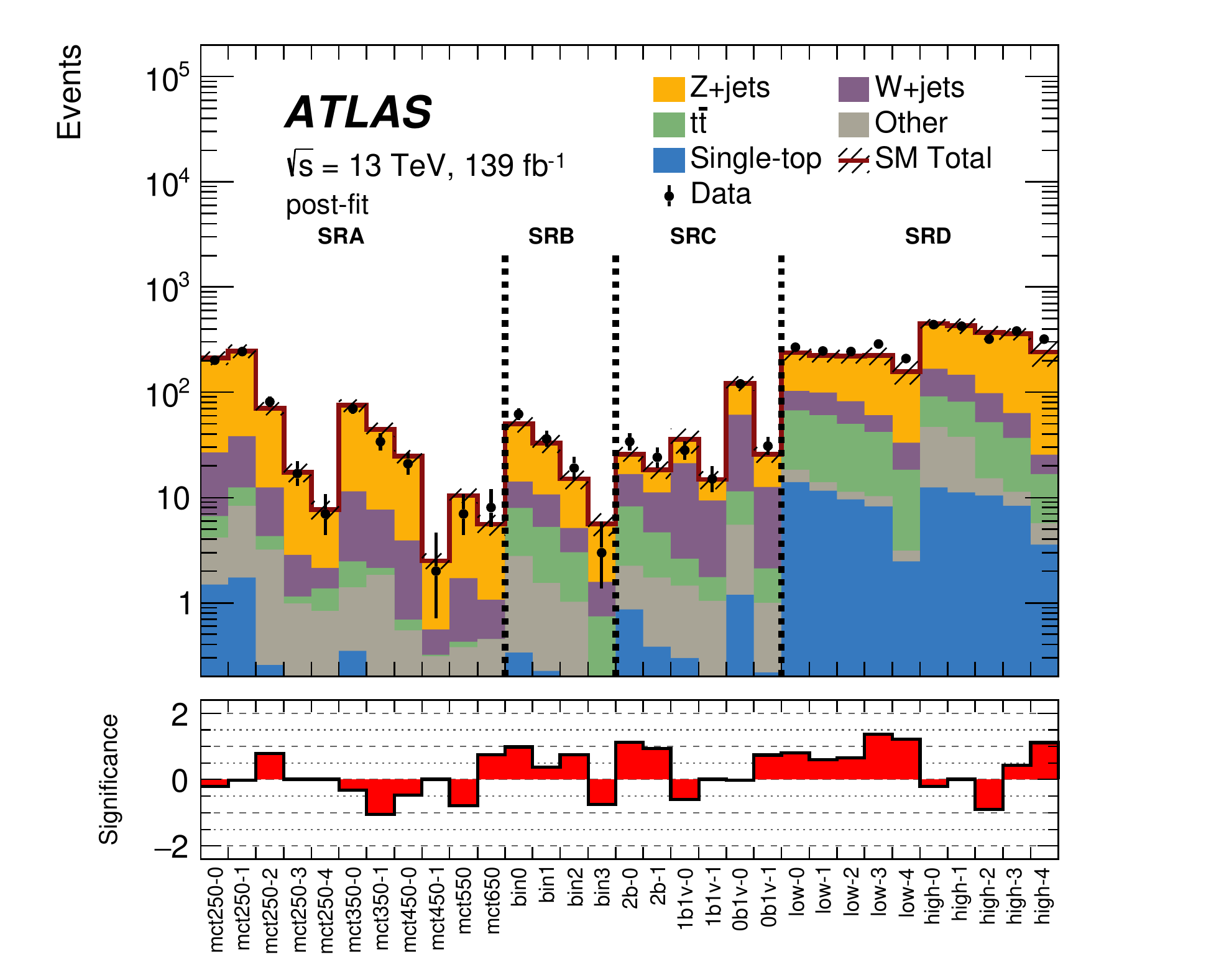}
\end{center}
\caption{Comparison between observed and background-only post-fit predicted background yields for the signal regions. The ratio plot represents  the statistical significance of the discrepancy between the observed and predicted value. The ``Other'' category includes contributions from diboson and $\ttbar+W/Z/H$ production. The shaded band represents the total uncertainty of the background prediction. Where present, the last digit in the signal region name labels the bins of \meff for \SRA{}, \XGB for \SRB, \met for \SRC and \cosbb for  \SRD.
}
\label{fig:pull_SR}
\end{figure}
 
\begin{figure}[!h]
\begin{center}
\subfigure[]{
\includegraphics[width=0.48\textwidth]{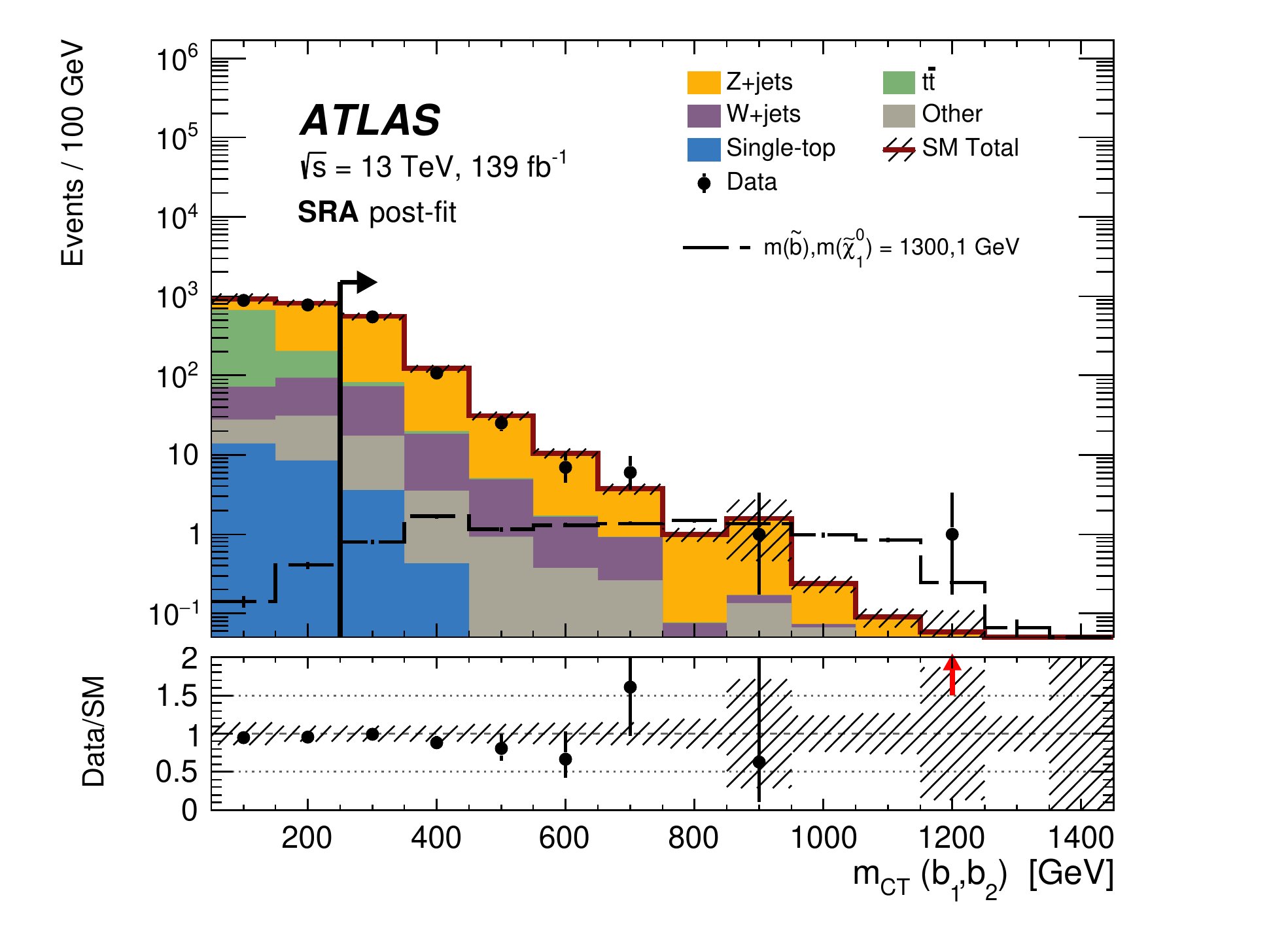}
}
\subfigure[]{
\includegraphics[width=0.48\textwidth]{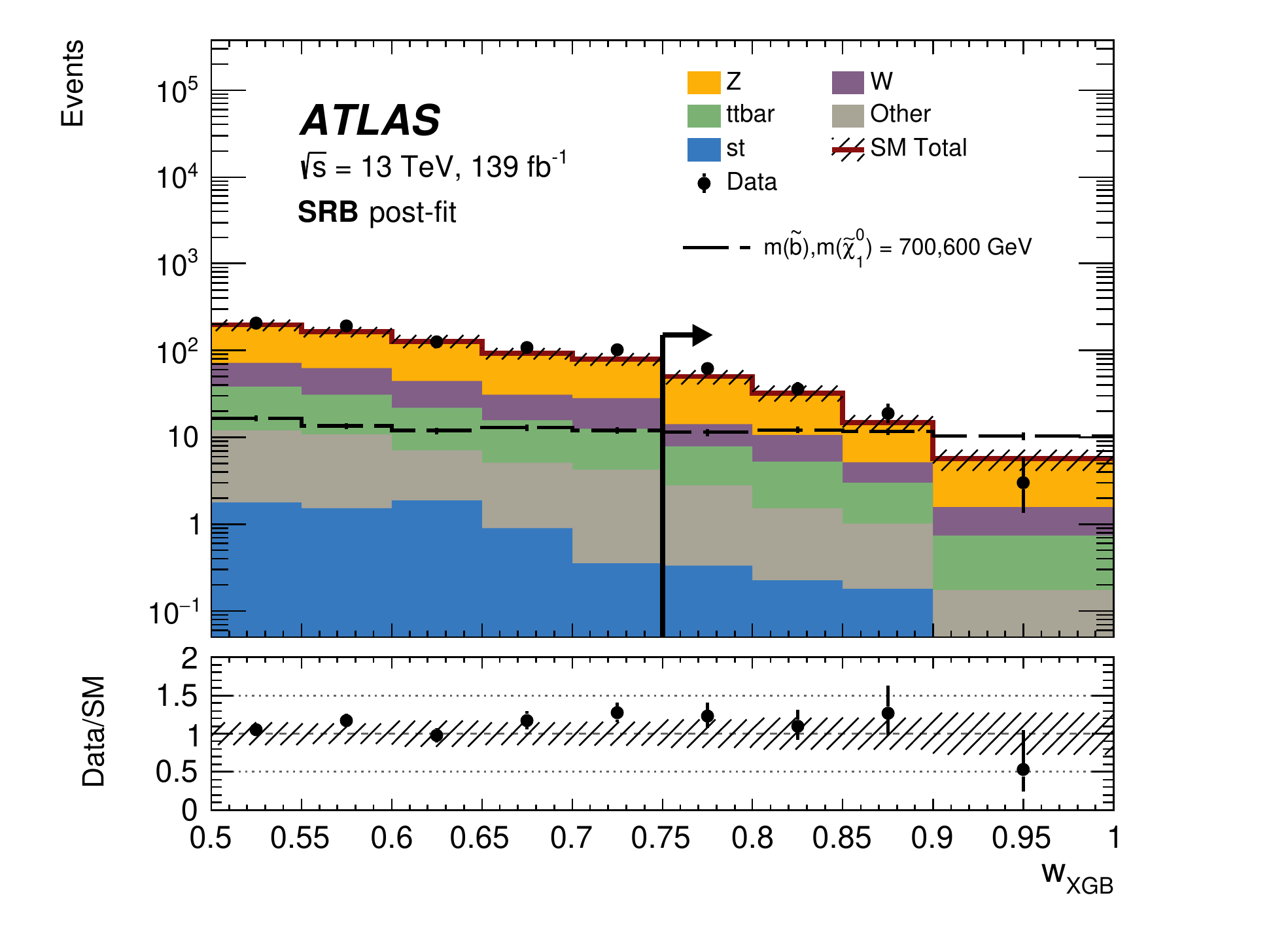}
}
\subfigure[]{
\includegraphics[width=0.48\textwidth]{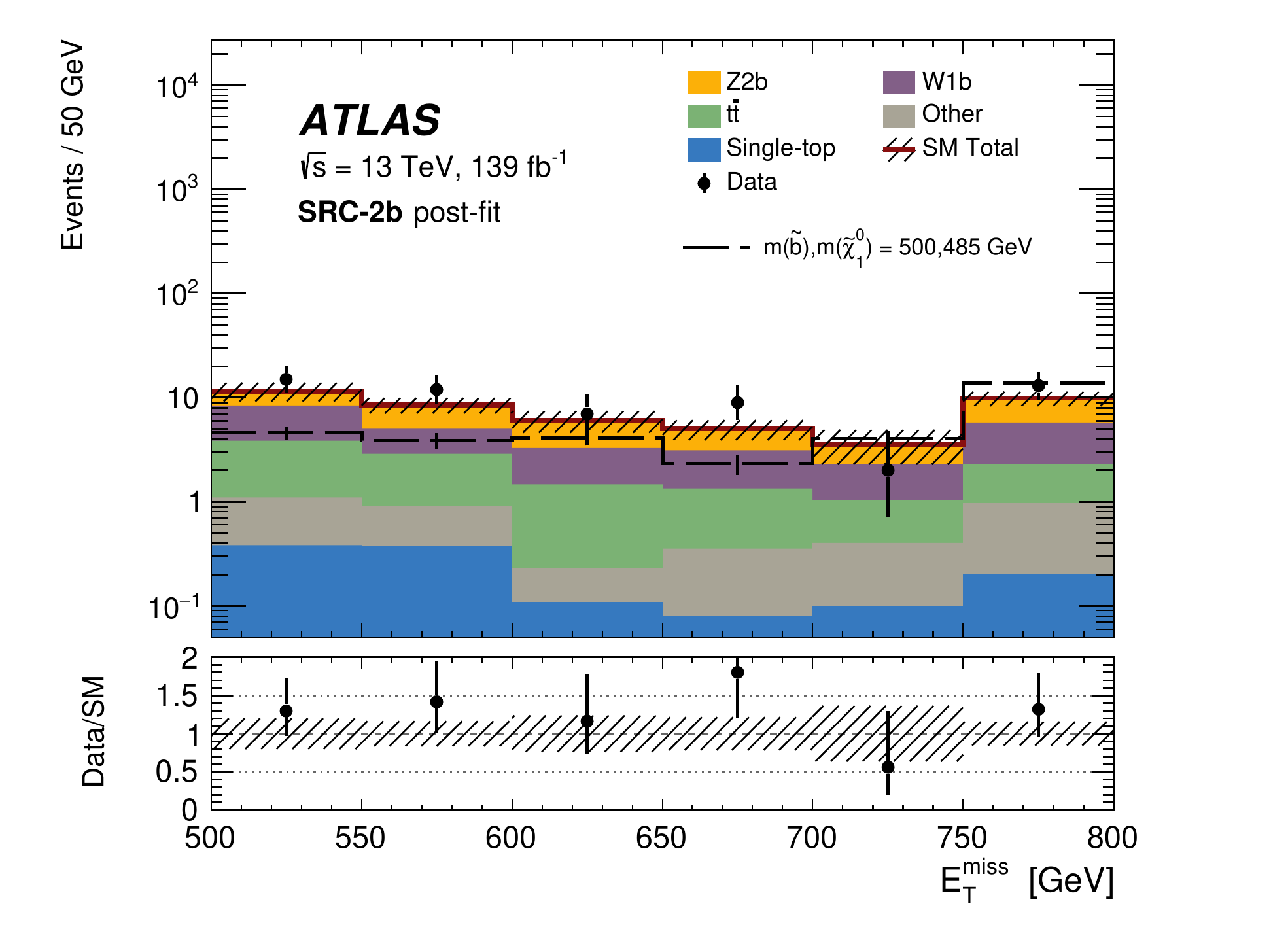}
}
\subfigure[]{
\includegraphics[width=0.48\textwidth]{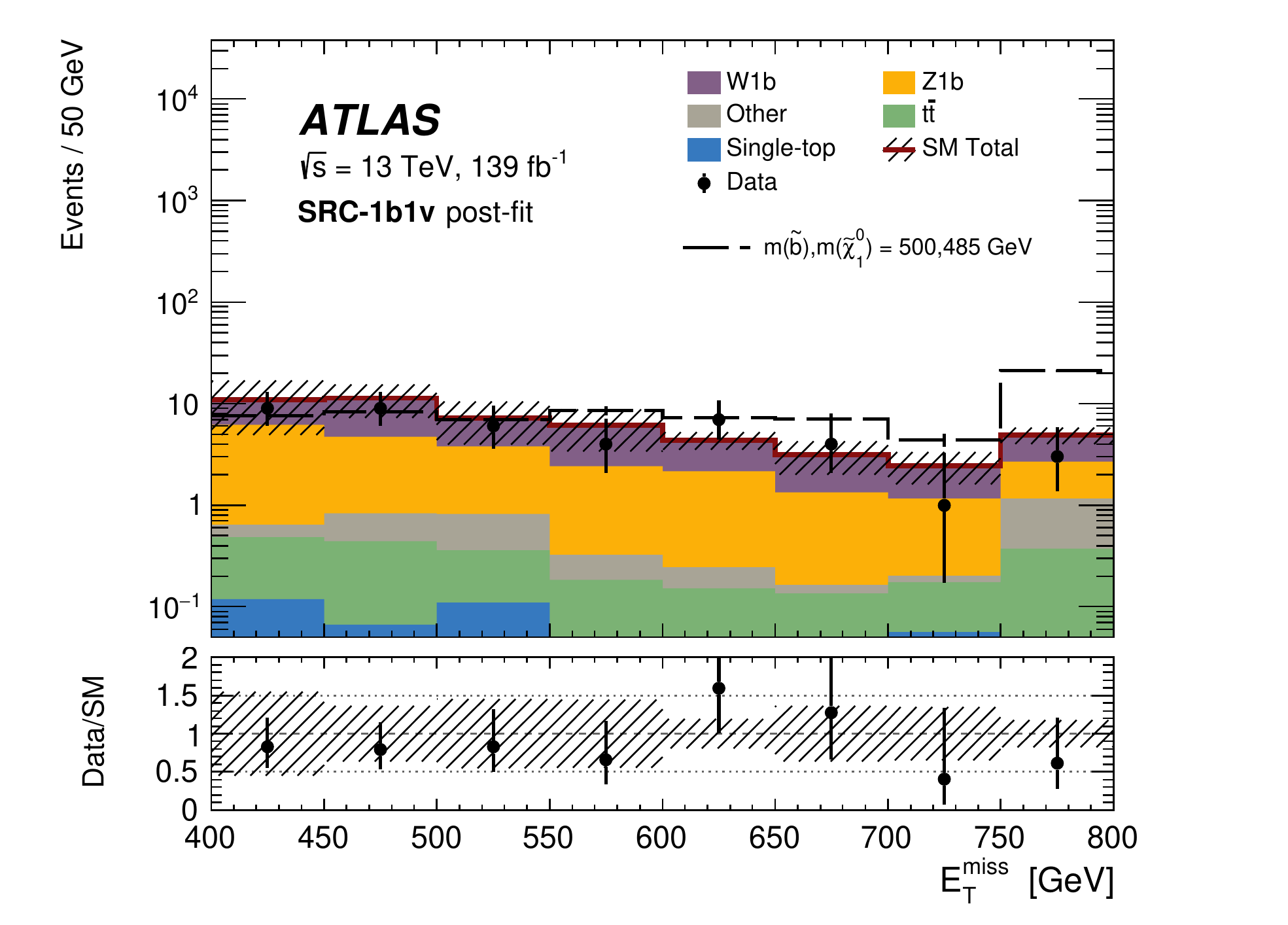}
}
\subfigure[]{
\includegraphics[width=0.48\textwidth]{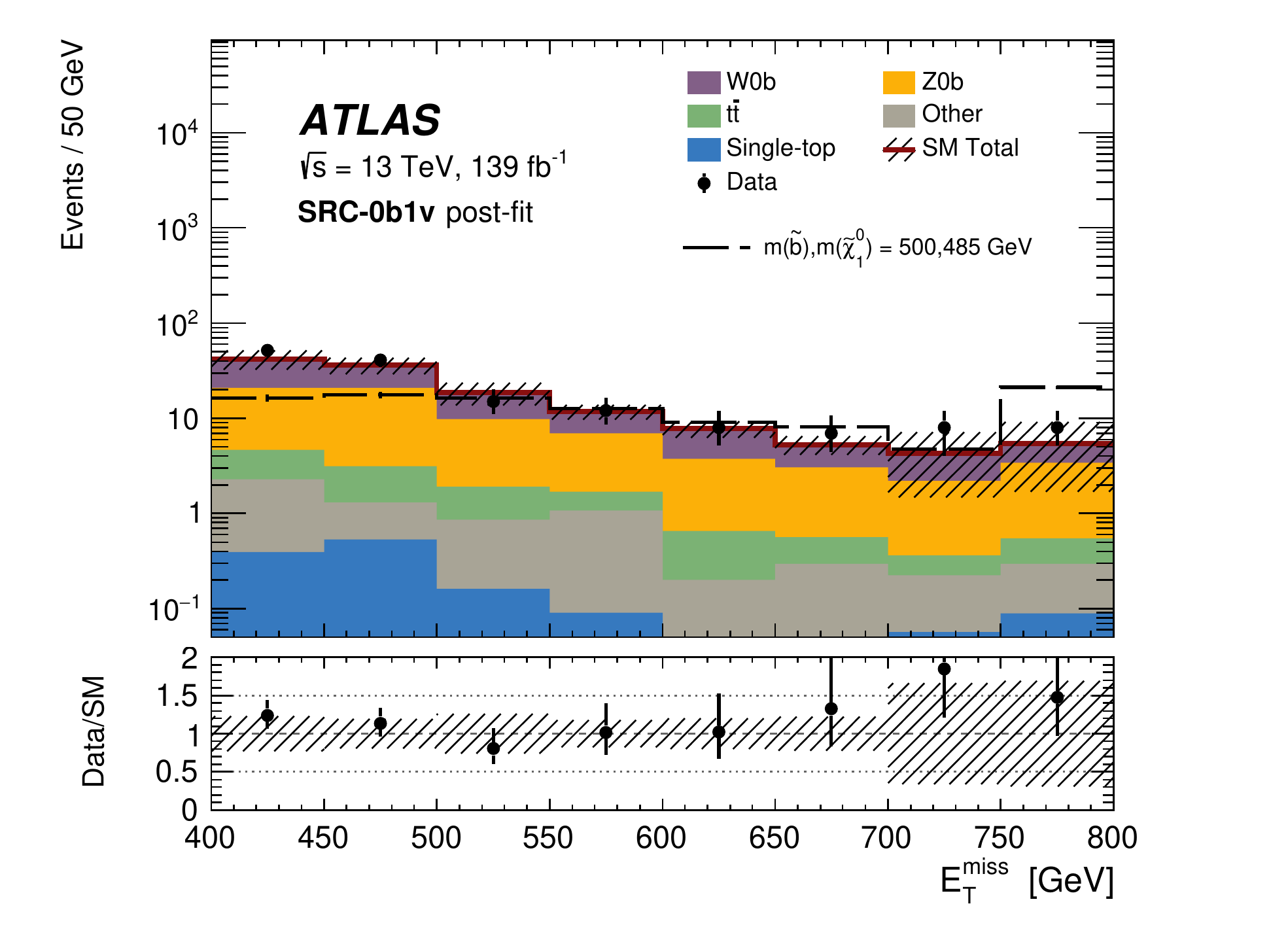}
}
\end{center}
\caption{Post-fit distribution of (a) \mct in \SRA; (b) \XGB in \SRB; and \met in (c) \SRCA, (d) \SRCB, (e) \SRCC. Where relevant, the signal region selection is indicated by a black arrow. In each of (a)--(e) the lower plot shows the ratios of the observed yields to the post-fit predicted background yields, with a red arrow indicating when a ratio value is outside the displayed interval. The ``Other'' category includes contributions from diboson, and $\ttbar+W/Z/H$ production. The shaded band represents the total uncertainty of the background prediction. The last bin includes overflow events.}
\label{fig:SR_N1s_1}
\end{figure}
 
\begin{figure}[!h]
\begin{center}
\subfigure[]{
\includegraphics[width=0.48\textwidth]{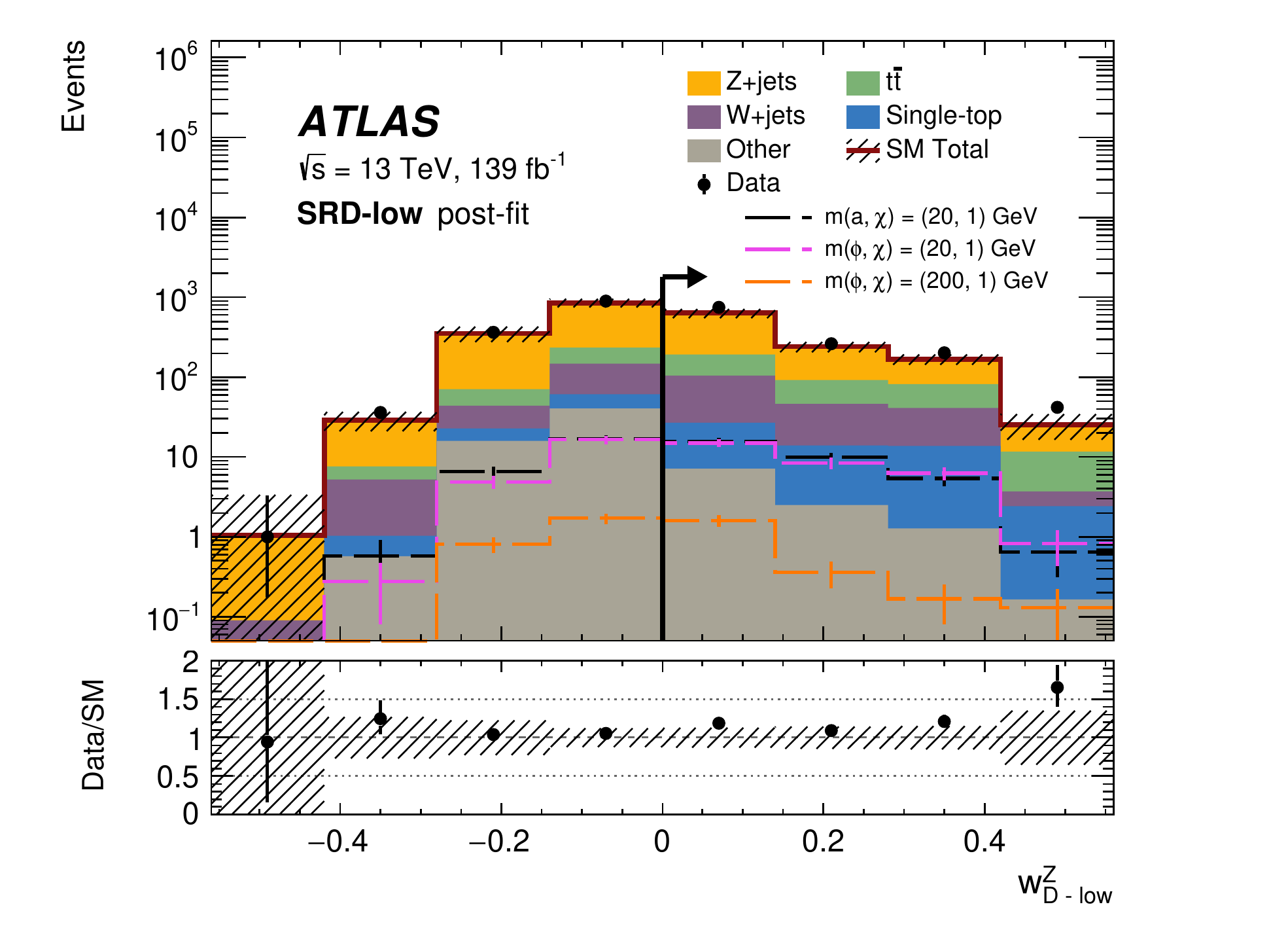}
}
\subfigure[]{
\includegraphics[width=0.48\textwidth]{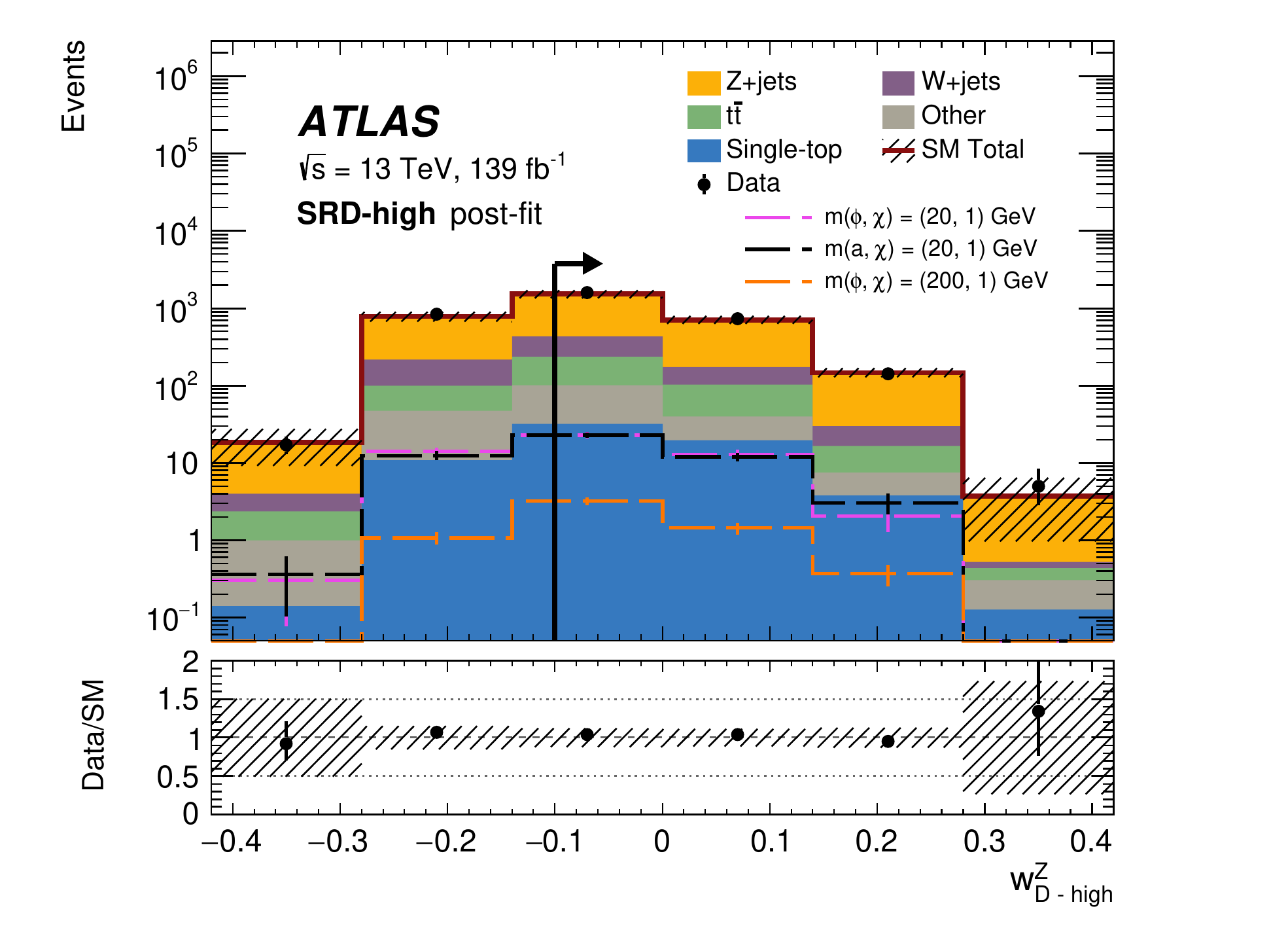}
}
\subfigure[]{
\includegraphics[width=0.48\textwidth]{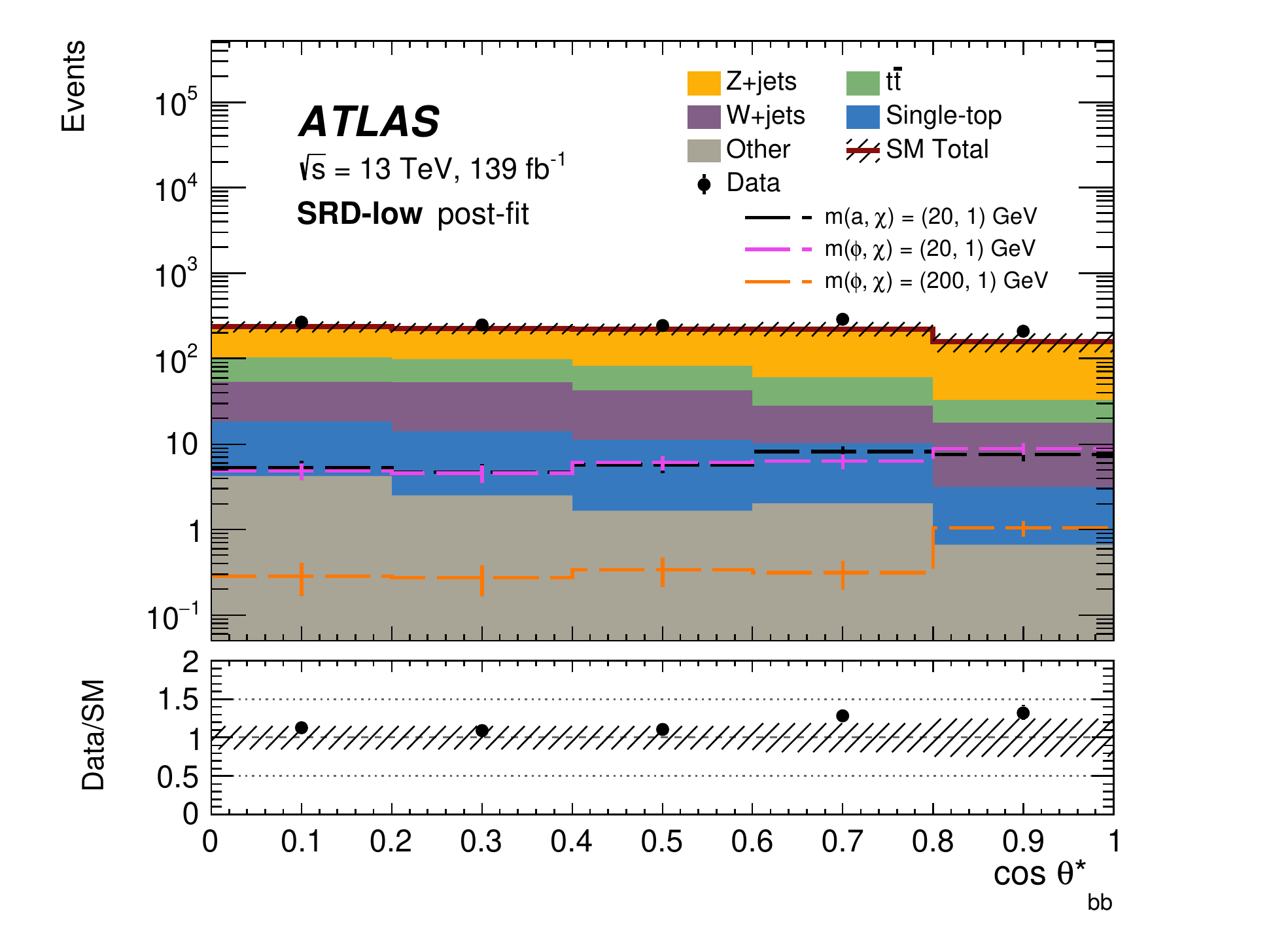}
}
\subfigure[]{
\includegraphics[width=0.48\textwidth]{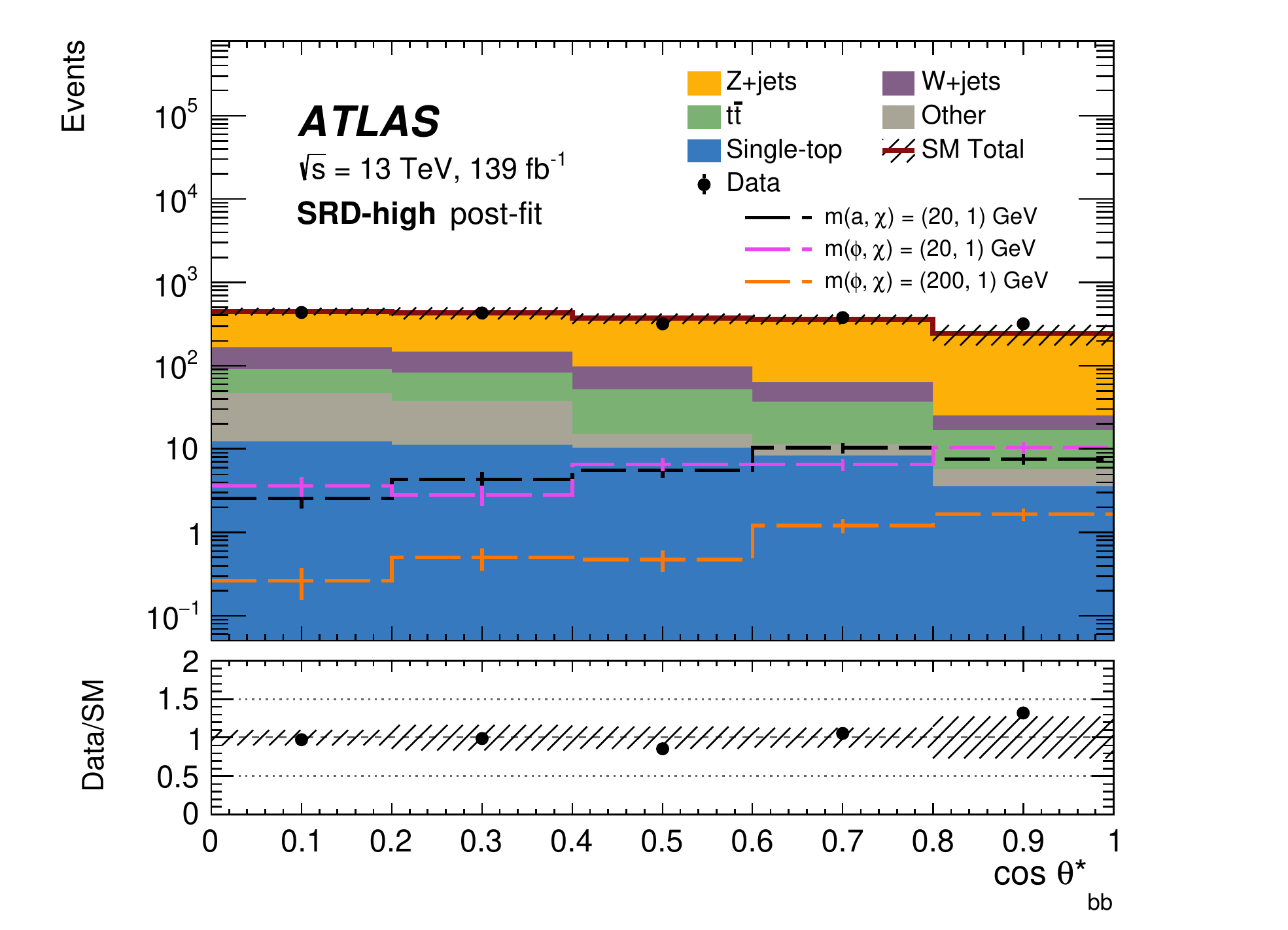}
}
\end{center}
\caption{Post-fit distribution of  (a) $\BDTZDA$ and (c) \cosbb in \SRDA and (b) $\BDTZDB$ and (d) \cosbb in \SRDB. Where relevant, the signal region selection is indicated by a black arrow. In each of (a)--(d) the lower plot shows the ratios of the observed yields to the post-fit predicted background yields. The ``Other'' category includes contributions from diboson and $\ttbar+W/Z/H$ production. The shaded band represents the total uncertainty of the background prediction. The last bin includes the overflow events.}
\label{fig:SR_N1s_2}
\end{figure}

The model-independent fit includes also the signal regions of Tables~\ref{tab:SRA_summary}, \ref{tab:SRB_summary}, \ref{tab:SRC_SR_summary} and \ref{tab:SRD_summary} in the likelihood as additional constraining bins. To maximise the model-independent sensitivity to new phenomena, no additional binning of \meff (for \SRA{}), \XGB (for \SRB), \met (\SRC) or \cosbb (in \SRD) is considered. In the case of \SRA{}, an additional selection on \mct is added with respect to Table~\ref{tab:SRA_summary}, corresponding to the  \SRA{mctXi} selections described in Section~\ref{sec:sbotA:SRs}. A profile-likelihood-ratio statistic~\cite{Cowan:2010js}, where a signal of intensity $\mu_{\mathrm{sig}}$ is assumed to contribute to the signal region only, is used to assess the $p$-value of the background-only hypothesis (corresponding to $\mu_{\mathrm{sig}} = 0$) and to extract $S^{95}_{\mathrm{exp}}$ and $S^{95}_{\mathrm{obs}}$, i.e.\ the expected and observed 95\% confidence level (CL) limits on $\mu_{\mathrm{sig}}$ using the \cls prescription. The limit $S^{95}_{\mathrm{obs}}$ is also converted to a limit on the visible cross-section $\sigma_{\mathrm{vis}}$ by dividing by the total integrated luminosity. The results are summarised in Table~\ref{tab:discovery_limits}.

\begin{table}[!htpb]
\caption{Left to right: SM expectation from background-only fit for the model-independent regions,  95\% CL upper limits on the visible cross-section ($\langle\epsilon\sigma\rangle_{\mathrm{obs}}^{95}$), on the observed ($S_{\mathrm{obs}}^{95}$ ) and expected ($S_{\mathrm{exp}}^{95}$) number of signal events. The last two columns
indicate the $CL_B$ value, i.e. the confidence level observed for the background-only hypothesis, and the discovery $p$-value ($p(s = 0)$) capped at a value of 0.5.}
\vspace{0.5cm}
\label{tab:discovery_limits}
\centering
\resizebox{\columnwidth}{!}{
\begin{tabular*}{\textwidth}{@{\extracolsep{\fill}}l|cc|ccccc}
\toprule
{Signal channel}          & Obs.            &  SM exp.  & $\langle\epsilon{\mathrm{\sigma}}\rangle_{\mathrm{obs}}^{95}$ [fb]  &  $S_{\mathrm{obs}}^{95}$  & $S_{\mathrm{exp}}^{95}$ & $CL_{B}$ & $p(s=0)$ ($Z$)  \\
\midrule
\SRA{mct250i}  & 552  & $555 \pm 75$ & $0.94$ &  $131$~~ & $ { 133 }^{ +47 }_{ -35 }$ & $0.48$ & $ 0.49$~$(0.03)$ \\%
\SRA{mct350i}  & 104  & $120 \pm 16$ & $0.17$ &  $24$ & ~$ { 32 }^{ +8 }_{ -9 }$ & $0.19$ & $ 0.5$~~~$(0)$~~~~~ \\%
\SRA{mct450i}  & ~~23 & $27.1 \pm 3.8$ & $0.06$ &  $8.7$ & $ { 12.3 }^{ +5.5 }_{ -3.7 }$ & $0.17$ & $ 0.5$~~~$(0)$~~~~~ \\%
\SRA{mct550i}  & ~~~~7  & $10.4 \pm 1.7$ & $0.04$ & $5.6$ & $ { 8.1 }^{ +3.9 }_{ -2.3 }$ & $0.14$ & $ 0.5$~~~$(0)$~~~~~ \\%
\SRA{mct650i}  & ~~~~8  & ~~$5.6 \pm 1.4 $ & $0.06$ & $8.5$ & $ { 6.7 }^{ +3.4 }_{ -2.0 }$ & $0.73$ & $ 0.24$~$(0.72)$ \\%
\SRB                 & ~~22  &$20.6 \pm 4.6$ & $0.11$ &  $15.3$ & $ { 14.8 }^{ +5.2 }_{ -3.2 }$ & $0.54$ & $ 0.40$~$(0.26)$ \\%
\SRCA              & ~~58  &  $44.4\pm 5.8$& $0.22$ &  $30.3$ & $ { 20.7 }^{ +8.1 }_{ -5.6 }$ & $0.88$ & $ 0.09$~$(1.33)$ \\%
\SRCB              & ~~43  & ~~$51\pm10$ & $0.13$ &  $17.6$ & $ { 21.2 }^{ +8.2 }_{ -5.8 }$ & $0.28$ & $ 0.5$~~~$(0)$~~~~~ \\%
\SRCC              & 151  & $148\pm 25$ & $0.37$ &  $51$ & $ { 50}^{ +18 }_{ -13 }$ & $0.54$ & $ 0.48$~$(0.2)$~~ \\%
\SRDA              & 497  & $381 \pm 76$& $1.8$~~ &  $250$& $ { 155 }^{ +65 }_{ -60 }$ & $0.91$ & $ 0.07$~$(1.48)$ \\%
\SRDB              & 320  & $242 \pm 66$  & $1.4$~~ &  $195$ & $ { 140 }^{ +48 }_{ -44 }$ & $0.82$ & $ 0.13$~$(1.13)$ \\%
\bottomrule
\end{tabular*}

}
\end{table}

Finally, the model-dependent fit includes the contributions of specific BSM models to all control and signal regions and considers the full binning defined in Sections~\ref{sec:sbotA:SRs}, \ref{sec:sbotB:SRs}, \ref{sec:sbotC:SRs} and \ref{sec:DM:SRs}, to determine 95\% CL exclusion limits on BSM particle masses and, where relevant, couplings.
 
The exclusion reach for the bottom squark pair production model is obtained by comparing, for each value of the bottom squark and neutralino masses, the expected \cls value of the combined \SRC fit and that of  a combined fit to \SRA{} and \SRB: the one yielding the smaller expected \cls value is chosen to evaluate the expected and observed \cls.   The result is shown in Figure~\ref{fig:sbottom_limit}. A deviation of the observed limit from the expected one is observed at $\left(m_{\sbottomone},m_{\ninoone}\right)$ of about $\left(1150~\GeV,700~\GeV\right)$ in Figure~\ref{fig:sbottom_limit}(a): it is due to a small deficit of events relative to the background prediction in \SRA{mct350},   \SRA{mct450} and  \SRA{mct550}, where the signal contributions from models in this region of the parameter space is found to be significant. The small excess of observed events in \SRCA is instead responsible for the weaker than expected limit for a broad range of values around $m_{\sbottomone} = 700~\GeV$ and $m_{\ninoone} > 20~\GeV$ in Figure~\ref{fig:sbottom_limit}(b). Finally, the transition between regions where \SRA, \SRB or \SRC dominate the sensitivity is responsible for the wiggles in the expected and observed limits for $650~\GeV < m_{\sbottomone} < 850~\GeV$ in Figure~\ref{fig:sbottom_limit}(a).
 
Bottom squark masses up to \limitLargeSplitSbottom are excluded for massless \ninoone. \SRC dominates the expected sensitivity up to $\dmBN \sim 70~\GeV$. The use of the \softbvrt algorithm largely contributed to the sensitivity for small \dmBN, allowing the exclusion of bottom squark masses up to \limitSmallSplitSbottom for  $\dmBN = \SplitLimit$: for example, the minimum  expected exclusion value of \dmBN for $m_{\sbottomone} = 600~\GeV$ using \SRCA alone would be about 25~\GeV.
 
\begin{figure}[!h]
\begin{center}
\subfigure[]{
\includegraphics[width=0.48\textwidth]{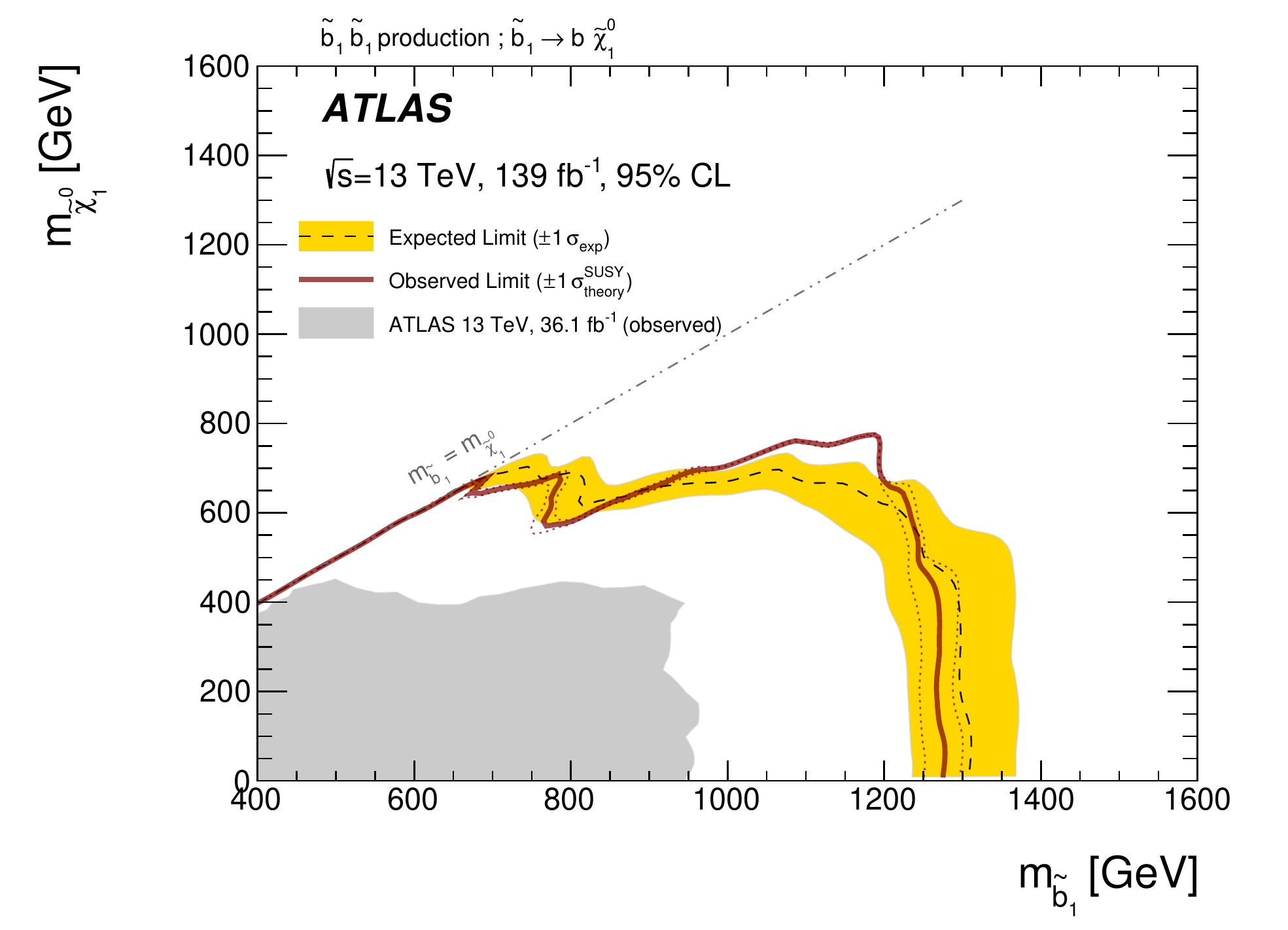}
}
\subfigure[]{
\includegraphics[width=0.48\textwidth]{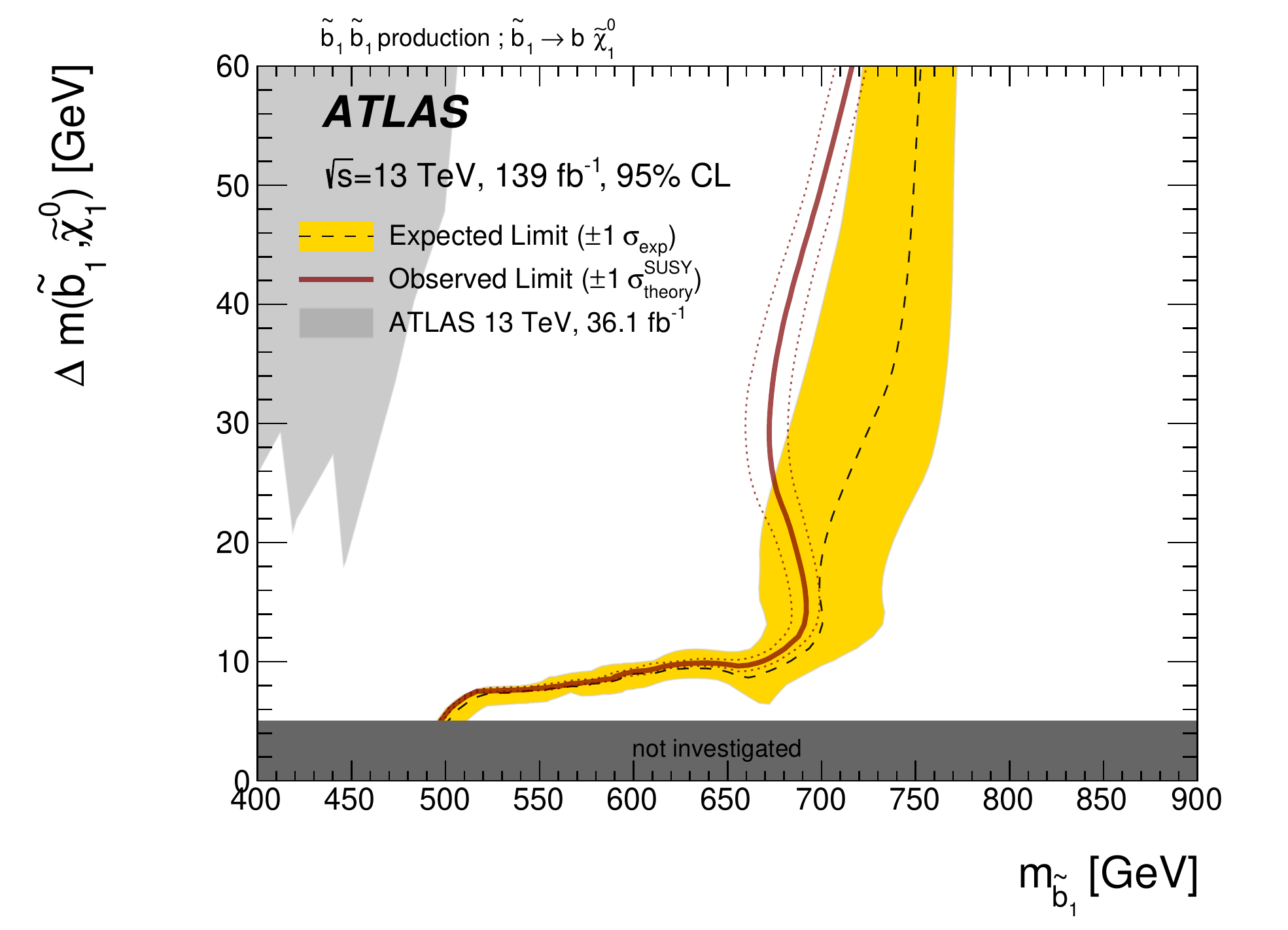}
}
\end{center}
\caption{Exclusion limits at 95\% CL on the masses of the \sbottomone and \ninoone displayed in the (a) $\left(m_{\sbottomone}, m_{\ninoone}\right)$ and (b) $\left(m_{\sbottomone}, \dmBN\right)$ planes. The dashed line and the shaded band are the expected limit and its $\pm1\sigma$ uncertainty, respectively. The thick solid line is the observed limit for the central value of the signal cross-section. The expected and observed limits do not include the effect of the theoretical uncertainties in the signal cross-section. The dotted lines show the effect on the observed limit of varying the signal cross-section by $\pm1\sigma$ of the theoretical uncertainty. Regions excluded by previous analyses are shaded in light grey, while the region with $\dmBN < 5~\GeV$, not investigated in this analysis, is shaded in dark grey.
}
\label{fig:sbottom_limit}
\end{figure}
 
The combined fit to \SRA{} and \SRB is also used to set limits on the model where pairs of  scalar third-generation down-type leptoquarks are produced. Limits at 95\% CL are set on the branching ratio $\lqdbr= 1-\mathcal{B}\left(\lqd\rightarrow b\nu_{\tau}\right)$ as a function of the leptoquark mass.   The result is shown in Figure~\ref{fig:lq_limit}. \lqd masses up to \limitLQ are excluded for $\lqdbr = 0$, while the limit decreases to 400~\GeV\ if $\lqdbr = 95\%$. The same event deficit in \SRA{mct350},   \SRA{mct450} and  \SRA{mct550} already discussed in relation to Figure~\ref{fig:sbottom_limit}(a) is responsible for the discrepancy between the observed and expected limit for leptoquark masses around  $700\ \GeV$.
 
\begin{figure}[!h]
\begin{center}
\includegraphics[width=0.48\textwidth]{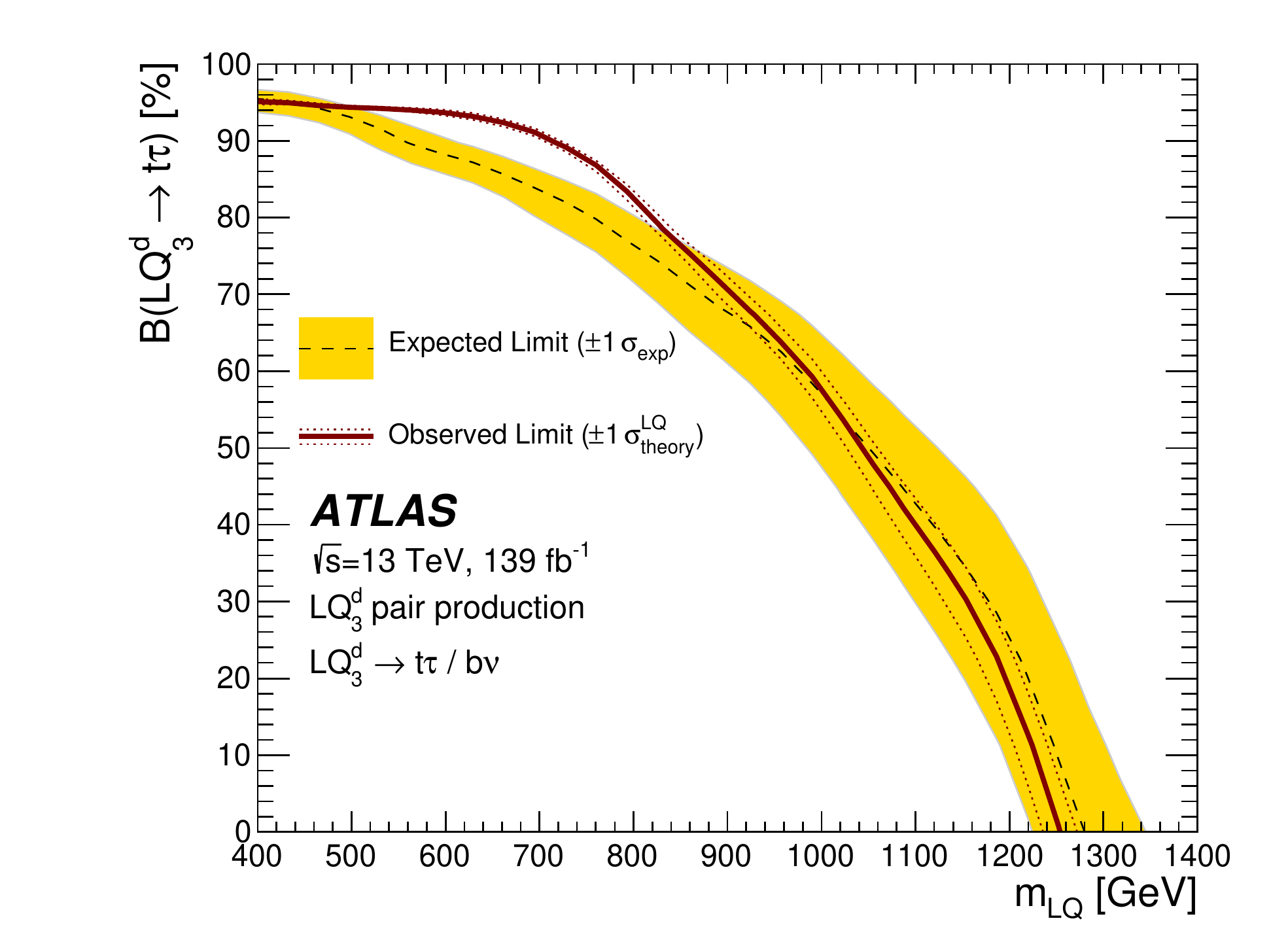}
\end{center}
\caption{Exclusion limits at 95\% CL on \lqdbr as a function of the \lqd mass for a model of \lqd pair production. The dashed line and the shaded band are the expected limit and its $\pm1\sigma$ uncertainty, respectively. The thick solid line is the observed limit for the central value of the signal cross-section. }
\label{fig:lq_limit}
\end{figure}
 
Finally, exclusion limits are set on DM production models with scalar or pseudoscalar mediators in Figure~\ref{fig:dm_limit}. The limits are obtained by taking for each point the signal region (either \SRDA or \SRDB) that yields the better expected \cls value. Limits on the cross-section are provided for a $\chi$ mass of 1 GeV and unity coupling of the mediator to the dark matter particles, separately for scenarios with a scalar or a pseudoscalar mediator. In both cases, cross-sections exceeding between 5 and 300  times the predicted rate for mediators with masses between 10  and 500~\GeV\ are excluded.

\begin{figure}[!h]
\begin{center}
\subfigure[]{
\includegraphics[width=0.48\textwidth]{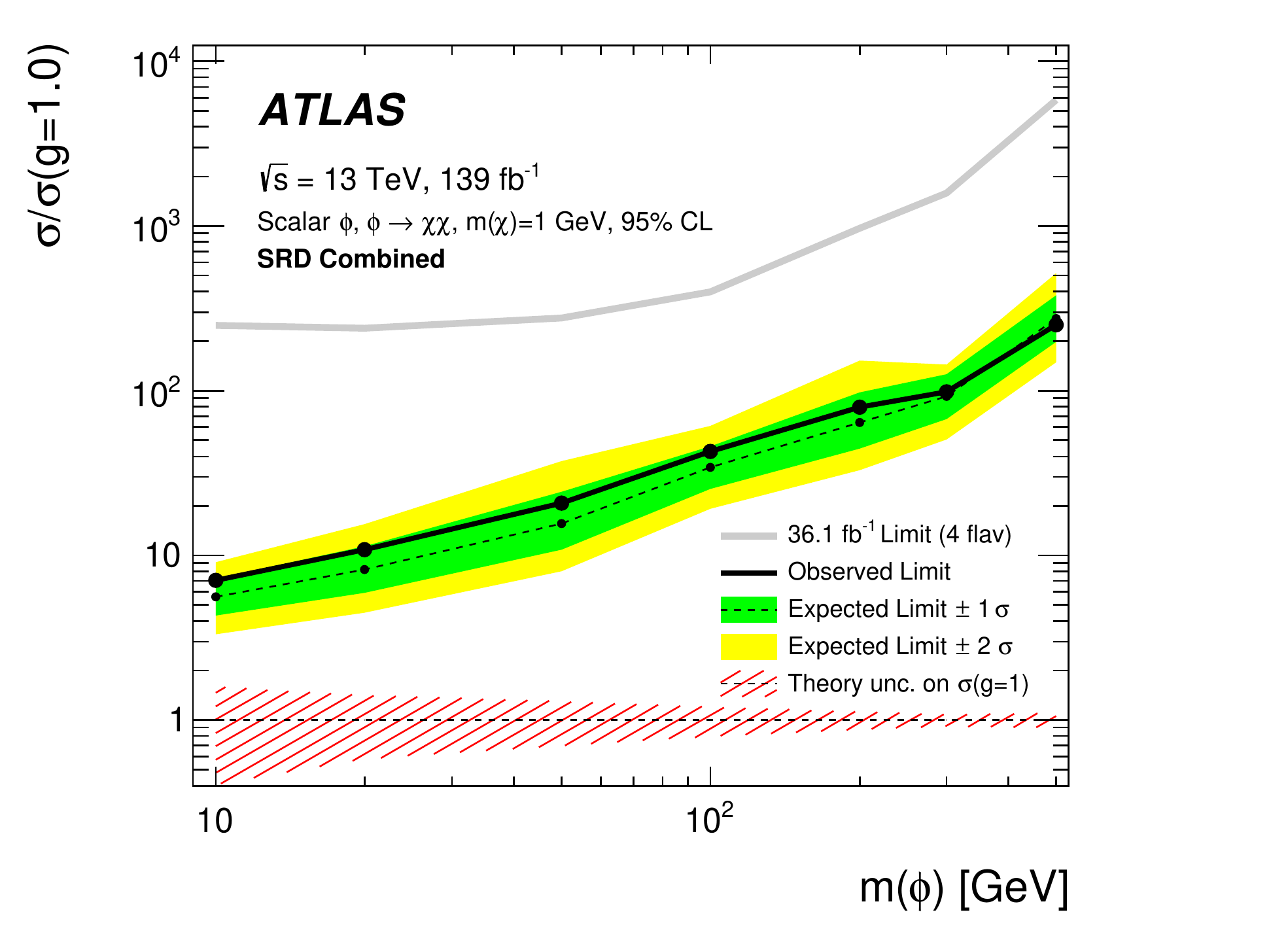}
}
\subfigure[]{
\includegraphics[width=0.48\textwidth]{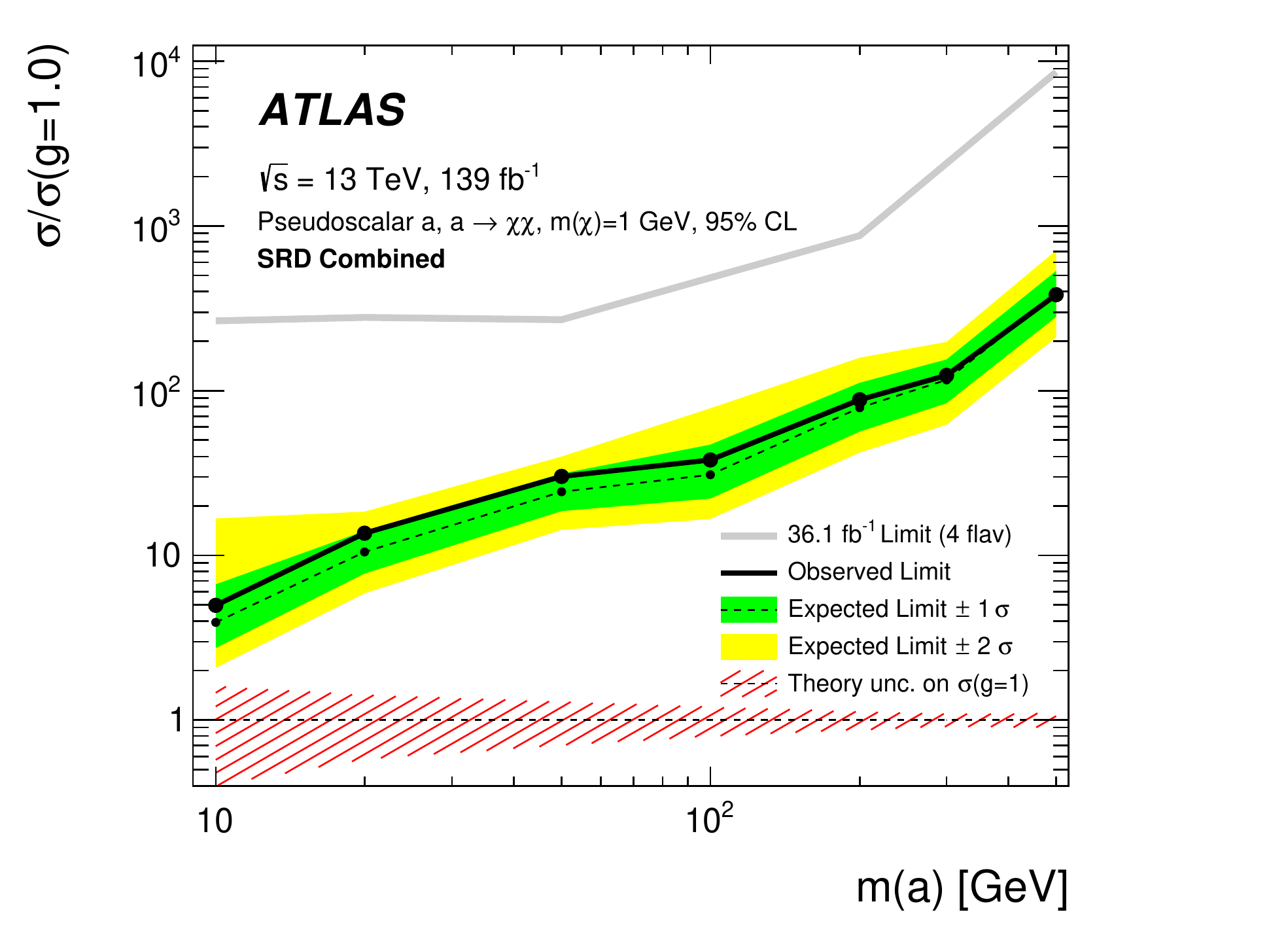}
}
\end{center}
\caption{95\% CL limits on the cross-section for the models of dark matter production mediated by a (a) scalar or (b) pseudoscalar mediator.  The dashed line and the shaded band are the expected limit and its $\pm1\sigma$ and $\pm2\sigma$ uncertainties, respectively. The thick solid line is the observed limit for the central value of the signal cross-section. The grey line shows the limit from previous results~\cite{SUSY-2016-18}. The hatched red band shows the uncertainty of the model cross-section.
}
\label{fig:dm_limit}
\end{figure}

 
\FloatBarrier
 
\section{Conclusions}
\label{sec:conclusion}
 
The result of a search for an excess of events containing large missing transverse momentum and at least one jet or decay vertex originating from the hadronisation and fragmentation of a $b$-quark is reported. The analysis uses $139$~fb$^{-1}$ of $pp$
collisions at $\sqrt{s}=13$~\TeV\ collected by the ATLAS experiment at the LHC between 2015 and 2018. No excess above the Standard Model prediction is observed for any of the signal regions considered. Limits on the visible cross-section of a generic BSM contribution in the signal regions have been set and range from 0.04 fb to 1.8 fb, depending on the signal region. The results are also used to set exclusion limits on specific BSM scenarios. In $R$-parity-conserving SUSY scenarios where bottom squarks decay into a $b$-quark and the lightest neutralino, the results are used to derive 95\% CL exclusion limits as a function of the \sbottomone and \ninoone masses. Bottom squark masses up to \limitLargeSplitSbottom are excluded for massless \ninoone, while the use of dedicated techniques to identify the secondary vertex from a low-\pt $b$-hadron allows the exclusion of bottom squark masses up to \limitSmallSplitSbottom for  $\dmBN = \SplitLimit$. The search is also interpreted to exclude pair-produced \lqd with masses up to \limitLQ (400~\GeV) for $\lqdbr = 0$ ($95\%$).
In scenarios where dark matter particles with a mass of $1~\GeV$ are produced via a scalar or pseudoscalar mediator in association with $b$-quarks, cross-sections exceeding between 5 and 300  times the predicted rate for mediators with masses between 10  and 500~\GeV\ and unitary coupling to the dark matter particles are excluded. All these limits extend substantially beyond the region of parameter space excluded by similar searches with the ATLAS detector using $36.1~\ifb$ of $pp$ collisions. This is due to several improvements in the analysis in addition to the increase in integrated luminosity; in particular the use of multivariate techniques and the TC-LVT algorithm which provide enhanced sensitivity to models with small \dmBN.

\FloatBarrier

\section*{Acknowledgements}
 

We thank CERN for the very successful operation of the LHC, as well as the
support staff from our institutions without whom ATLAS could not be
operated efficiently.
 
We acknowledge the support of ANPCyT, Argentina; YerPhI, Armenia; ARC, Australia; BMWFW and FWF, Austria; ANAS, Azerbaijan; SSTC, Belarus; CNPq and FAPESP, Brazil; NSERC, NRC and CFI, Canada; CERN; ANID, Chile; CAS, MOST and NSFC, China; Minciencias, Colombia; MSMT CR, MPO CR and VSC CR, Czech Republic; DNRF and DNSRC, Denmark; IN2P3-CNRS and CEA-DRF/IRFU, France; SRNSFG, Georgia; BMBF, HGF and MPG, Germany; GSRT, Greece; RGC and Hong Kong SAR, China; ISF and Benoziyo Center, Israel; INFN, Italy; MEXT and JSPS, Japan; CNRST, Morocco; NWO, Netherlands; RCN, Norway; MNiSW and NCN, Poland; FCT, Portugal; MNE/IFA, Romania; JINR; MES of Russia and NRC KI, Russian Federation; MESTD, Serbia; MSSR, Slovakia; ARRS and MIZ\v{S}, Slovenia; DST/NRF, South Africa; MICINN, Spain; SRC and Wallenberg Foundation, Sweden; SERI, SNSF and Cantons of Bern and Geneva, Switzerland; MOST, Taiwan; TAEK, Turkey; STFC, United Kingdom; DOE and NSF, United States of America. In addition, individual groups and members have received support from BCKDF, CANARIE, Compute Canada, CRC and IVADO, Canada; Beijing Municipal Science \& Technology Commission, China; COST, ERC, ERDF, Horizon 2020 and Marie Sk{\l}odowska-Curie Actions, European Union; Investissements d'Avenir Labex, Investissements d'Avenir Idex and ANR, France; DFG and AvH Foundation, Germany; Herakleitos, Thales and Aristeia programmes co-financed by EU-ESF and the Greek NSRF, Greece; BSF-NSF and GIF, Israel; La Caixa Banking Foundation, CERCA Programme Generalitat de Catalunya and PROMETEO and GenT Programmes Generalitat Valenciana, Spain; G\"{o}ran Gustafssons Stiftelse, Sweden; The Royal Society and Leverhulme Trust, United Kingdom.
 
The crucial computing support from all WLCG partners is acknowledged gratefully, in particular from CERN, the ATLAS Tier-1 facilities at TRIUMF (Canada), NDGF (Denmark, Norway, Sweden), CC-IN2P3 (France), KIT/GridKA (Germany), INFN-CNAF (Italy), NL-T1 (Netherlands), PIC (Spain), ASGC (Taiwan), RAL (UK) and BNL (USA), the Tier-2 facilities worldwide and large non-WLCG resource providers. Major contributors of computing resources are listed in Ref.~\cite{ATL-SOFT-PUB-2020-001}.
 

\printbibliography
\newpage

\clearpage 
 
\begin{flushleft}
\hypersetup{urlcolor=black}
{\Large The ATLAS Collaboration}

\bigskip

\AtlasOrcid[0000-0002-6665-4934]{G.~Aad}$^\textrm{\scriptsize 101}$,    
\AtlasOrcid[0000-0002-5888-2734]{B.~Abbott}$^\textrm{\scriptsize 127}$,    
\AtlasOrcid[0000-0002-7248-3203]{D.C.~Abbott}$^\textrm{\scriptsize 102}$,    
\AtlasOrcid[0000-0002-2788-3822]{A.~Abed~Abud}$^\textrm{\scriptsize 36}$,    
\AtlasOrcid[0000-0002-1002-1652]{K.~Abeling}$^\textrm{\scriptsize 53}$,    
\AtlasOrcid[0000-0002-2987-4006]{D.K.~Abhayasinghe}$^\textrm{\scriptsize 93}$,    
\AtlasOrcid[0000-0002-8496-9294]{S.H.~Abidi}$^\textrm{\scriptsize 29}$,    
\AtlasOrcid[0000-0002-8279-9324]{O.S.~AbouZeid}$^\textrm{\scriptsize 40}$,    
\AtlasOrcid{N.L.~Abraham}$^\textrm{\scriptsize 155}$,    
\AtlasOrcid[0000-0001-5329-6640]{H.~Abramowicz}$^\textrm{\scriptsize 160}$
% ...
\bigskip
\\

% ...
$^{29}$Physics Department, Brookhaven National Laboratory, Upton NY; United States of America.\\
% ...
$^{36}$CERN, Geneva; Switzerland.\\
% ...
$^{40}$Niels Bohr Institute, University of Copenhagen, Copenhagen; Denmark.\\
% ...
$^{53}$II. Physikalisches Institut, Georg-August-Universit\"{a}t G\"ottingen, G\"ottingen; Germany.\\
% ...
$^{93}$Department of Physics, Royal Holloway University of London, Egham; United Kingdom.\\
% ...
$^{101}$CPPM, Aix-Marseille Universit\'e, CNRS/IN2P3, Marseille; France.\\
$^{102}$Department of Physics, University of Massachusetts, Amherst MA; United States of America.\\
% ...
$^{127}$Homer L. Dodge Department of Physics and Astronomy, University of Oklahoma, Norman OK; United States of America.\\
% ...
$^{155}$Department of Physics and Astronomy, University of Sussex, Brighton; United Kingdom.\\
% ...
$^{160}$Raymond and Beverly Sackler School of Physics and Astronomy, Tel Aviv University, Tel Aviv; Israel.\\
% ...

\end{flushleft}


\end{document}